\documentclass[final,5p,times,twocolumn,numbers,sort&compress]{elsarticle}

\usepackage{graphicx}
\usepackage{amsmath}
\usepackage{subfigure}
\usepackage{wrapfig}
\usepackage{epsfig}

\newcommand\fverb{\setbox\fverbbox=\hbox\bgroup\verb}
\newcommand\fverbdo{\egroup\medskip\noindent%
			\fbox{\unhbox\fverbbox}\ }
\newcommand\fverbit{\egroup\item[\fbox{\unhbox\fverbbox}]}
\newbox\fverbbox

\newcommand{\be}{\begin{equation}}
\newcommand{\ee}{\end{equation}}
\newcommand{\bea}{\begin{eqnarray}}
\newcommand{\eea}{\end{eqnarray}}
\newcommand{\nn}{\nonumber}

\def\half{{\textstyle{1\over2}}}

\def\CL{{\cal L}}

\def\CV{{\cal V}}

\def\Tr{{\sf Tr}}

\def\bar{\overline}

\def\tilde{\widetilde}

\def\half{{\scriptstyle \raise.15ex\hbox{${1\over2}$}}}

\newcommand{\beq}{\begin{equation}}
\newcommand{\eeq}{\end{equation}}

\newcommand{\real}{\relax{\rm I\kern-.18em R}}

\newcommand{\tr}{\mbox{\,tr\,}}

\newcommand{\eff}{{\rm eff}}
\newcommand{\diag}{{\rm diag}}

\newcommand{\Cite}[1]{$\,$\cite{#1}}

\def\vek#1{{\bf #1}}

\usepackage{graphicx}

\usepackage{afterpage,float}

\usepackage{amssymb}

\journal{Physics Letters B}

\begin{document}

\begin{frontmatter}



\title{Nearly conformal gauge theories in finite volume}
 
 \author[wupi]{Zoltan Fodor}
 
 \author[uop]{Kieran Holland}
 
 \author[ucsd]{Julius Kuti\corref{cor1}}
 \ead{jkuti@ucsd.edu}
 
 \author[ucsd]{D\'{a}niel N\'{o}gr\'{a}di}
 
 \author[ucsd]{Chris Schroeder}

 \cortext[cor1]{Corresponding author}
 
 \address[wupi]{Department of Physics, University of Wuppertal\\
Gaussstrasse 20, D-42119, Germany}
 \address[uop]{Department of Physics, University of the Pacific\\
3601 Pacific Ave, Stockton CA 95211, USA}
 \address[ucsd]{Department of Physics 0319, University of California, San Diego\\
9500 Gilman Drive, La Jolla, CA 92093, USA}



\begin{abstract}
We report new results on nearly conformal gauge theories with fermions in the fundamental 
representation of the SU(3) color gauge group as the number of  fermion flavors is varied 
in the $N_f =4-16$ range. To unambiguously
identify the chirally broken phase below the conformal window we apply a comprehensive
lattice tool set in finite volumes which includes the test of Goldstone pion dynamics, the
spectrum of the fermion Dirac operator, and eigenvalue distributions of random matrix theory.
We also discuss the theory inside the conformal window and present our 
first results on the running of the renormalized gauge coupling and the renormalization group beta function.
The importance of understanding finite volume zero momentum gauge field dynamics inside the conformal
window  is illustrated.  Staggered lattice fermions are used throughout the calculations.
\end{abstract}


\end{frontmatter}



\section{Introduction}
The Large Hadron Collider will probe the mechanism of electroweak symmetry breaking.
It is an intriguing possibility that new physics beyond
the Standard Model might take the form of some new strongly-interacting gauge
theory. In one scenario, the Higgs sector of the electroweak theory
is replaced by a so-called technicolor theory, whose dynamics provides
the required spontaneous symmetry breaking 
\cite{Weinberg:1979bn,Susskind:1978ms,Farhi:1980xs}. 
These models avoid the fine-tuning problem and may lead to a heavy 
composite Higgs particle on the TeV scale. 
Although attractive, the
challenge is to extend a technicolor theory to include fermion mass
generation, while satisfying the various constraints of electroweak
phenomenology. This  idea has lately been revived by new
explorations of the multi-dimensional theory space of nearly conformal gauge theories
\cite{Sannino:2009aw,Ryttov:2007sr,Dietrich:2006cm,Hong:2004td}. 
The terminology of {\em technicolor} in this report will refer in a generic sense to 
these investigations.
Exploring the new technicolor ideas has to be based on nonperturbative 
studies which are only becoming feasible now with the advent of new lattice technologies.

Model building of a strongly interacting electroweak sector
requires the knowledge of the phase diagram of nearly conformal gauge theories as the number of colors $N_c$, number of
fermion flavors $N_f$, and the fermion representation $R$ of the technicolor group are varied in theory space. 
For fixed $N_c$ and $R$ the theory is in the  
chirally broken phase for low $N_f$ and asymptotic freedom is maintained with a negative $\beta$ function.
On the other hand, if $N_f$ is large enough, the $\beta$
function is positive for all  couplings, and the theory is
trivial. If the regulator cut-off is removed, we are left with a free
non-interacting continuum theory.
There is some range
of $N_f$ for which the $\beta$ function might have a non-trivial zero, an
infrared fixed point, where the theory is in fact conformal 
\cite{Caswell:1974gg,Banks:1981nn}. This
method has been refined by estimating the critical value of $N_f$,
above which spontaneous chiral symmetry breaking no longer occurs 
\cite{Appelquist:1988yc,Cohen:1988sq,Appelquist:1996dq}. 

Interesting models require the theory  to be very
close to, but below, the conformal window, with a running coupling
which is almost constant over a large energy range \cite{Holdom:1981rm,Yamawaki:1985zg,Appelquist:1986an,Miransky:1996pd,Eichten:1979ah}. 
The nonperturbative knowledge of the critical $N^{crit}_f$ separating the two phases is essential and 
this has generated much interest and many new lattice
studies~\cite{Fodor:2008hn,Fodor:2008hm,Fodor:2009nh,DeGrand:2008kx,DeGrand:2008dh,
Svetitsky:2008bw,Shamir:2008pb,
Appelquist:2009ty,Appelquist:2007hu,Fleming:2008gy,DelDebbio:2008tv,DelDebbio:2008zf,DelDebbio:2008wb,
Hietanen:2009az,Hietanen:2008mr,Hietanen:2008vc,Deuzeman:2009mh,Deuzeman:2008da,Deuzeman:2008pf,
Deuzeman:2008sc,Catterall:2007yx,Catterall:2008qk,Jin:2008rc,Hasenfratz:2009ea,
DeGrand:2009mt,DeGrand:2009et,DelDebbio:2009fd}.

Our goal to unambiguously
identify the chirally broken phase below the conformal window requires the application and testing
of  a comprehensive lattice tool set in finite volumes which includes the test of Goldstone pion dynamics, the
spectrum of the fermion Dirac operator, and eigenvalue distributions of Random Matrix Theory (RMT).
Inside the conformal window we investigate the running coupling and the $\beta$ function. We report new
results at $N_f=4,8,9,12,16$ for fermions in the 
fundamental representation of the SU(3) technicolor gauge group. 
We find $N_f=4,8,9$  to be in the chirally broken phase and $N_f=16$ is consistent with the expected location
inside the conformal window. To resolve the $N_f=12$ phase from our simulations will require further analysis.

\section{Chiral symmetry breaking below the conformal window}

We will identify in lattice simulations the chirally broken phases with $N_f=4,8,9$ flavors of 
staggered fermions in the fundamental SU(3) color representation using finite volume analysis. 
The staggered fermions are deployed with a special 6-step exponential (stout) 
smearing procedure~\cite{Morningstar:2003gk} in the lattice action to reduce well-known cutoff effects with taste breaking 
in the Goldstone spectrum. The presence of taste breaking requires
a brief explanation of how staggered chiral perturbation theory
is applied in our analysis. The important work of Lee, Sharpe, Aubin and Bernard~\cite{Lee:1999zxa,Aubin:2003mg,Aubin:2003uc} 
is closely followed in the discussion.

\subsection{Staggered chiral perturbation theory}

Starting with the $N_f=4$ example~\cite{Lee:1999zxa}, the spontaneous breakdown of $SU(4)_L\times SU(4)_R $ to vector
$ SU(4) $ gives rise to 15 Goldstone modes, described by fields $\phi_i$.
These can be organized into an $SU(4)$ matrix
\begin{equation}
\Sigma(x) = \exp\Bigr(  i \frac{\phi }{ \sqrt{2}F}\Bigl)   \,,~~~~
\phi = \sum_{a=1}^{15} \phi_a T_a \,,
\end{equation}
where $F$ is the Goldstone decay constant in the chiral limit and the normalization 
$
T_a = \left\{ \xi_\mu, i\xi_{\mu5}, i\xi_{\mu\nu}, \xi_5 \right\} 
$
is used for the flavor generators.
The leading order chiral Lagrangian is given by
\begin{equation}
{\cal L}_\chi^{(4)} = \frac{F^2}{4} {\sf Tr} (\partial_\mu \Sigma 
\partial_{\mu} \Sigma^{\dagger} )  - 
\frac{1}{2} B\, m_q \, F^2 {\sf Tr}( \Sigma + \Sigma^{\dagger} ) \,,
\label{eq:Lagrange}
\end{equation}
with the fundamental parameters $F$ and $B$  measured on the technicolor scale $\Lambda_{\rm  TC}$ 
which replaced $\Lambda_{\rm  QCD}$ in the new theory.
Expanding the chiral Lagrangian in powers of $\phi$ one finds 15 degenerate pions
with masses given by
\begin{equation}
M_\pi^2 = 2 B m_q \left[1 + O(m_q/\Lambda_{\rm TC}) \right] \,.
\label{eq:mpisqchiral}
\end{equation}
The leading order term is the tree-level result,
while the corrections come from loop diagrams
and from higher order terms in the chiral Lagrangian.
The addition of $a^2 {\cal L}_\chi^{(6)}$ breaks chiral symmetry and 
lifts the degeneracy of the Goldstone pions. Correction terms are added to  Eq.~(\ref{eq:mpisqchiral})
which becomes 
\begin{equation}
M_\pi^2 = C(T_a)\cdot a^2\Lambda_{\rm TC}^4 +
2 B m_q \left[1 + O(m_q/\Lambda_{\rm TC}) + O(a^2\Lambda_{\rm TC}^2) \right] 
\label{eq:Mpi2}
\end{equation}
where the representation dependent $C(T_a)$ is a constant of order unity.
Contributions proportional to $a^2$ are due to ${\cal L}_\chi^{(6)} $,
and lead to massive Goldstone pions even in the $m_q\to 0$ chiral limit.
The only exception is the pion with flavor $\xi_5$ which remains 
massless because the $U(1)_A$ symmetry is protected.

Lee and Sharpe observe that the part of $\CL_\chi^{(6)}$
without derivatives, defining the potential $\CV_\chi^{(6)}$, is invariant
under flavor $SO(4)$ transformations and gives rise  to the $a^2$ term
in $M_\pi^2$.
Terms in $\CL_\chi^{(6)}$ involving derivatives break $SO(4)$ further down to 
the lattice symmetry group and give rise to 
non-leading terms proportional to $a^2 m$ and $a^4$.
The taste breaking potential is given by
\begin{eqnarray}
\hskip 0.2in - \CV_\chi^{(6)}  &= &
 C_1 \Tr(\xi_5\Sigma \xi_5 \Sigma^\dagger)
\nonumber \\ &&
+\, C_2\, \half 
\left[\Tr(\Sigma^2) -\Tr(\xi_5\Sigma\xi_5\Sigma) + h.c.\right] \,
\nonumber \\ &&
+\, C_3\, \half \sum_\nu \left[\Tr(\xi_\nu\Sigma\xi_\nu \Sigma) + h.c.\right] 
\nonumber \\ &&
+\, C_4\, \half \sum_\nu 
\left[\Tr(\xi_{\nu5}\Sigma\xi_{5\nu} \Sigma) + h.c.\right] 
\nonumber \\ &&
+\, C_5\, \half \sum_\nu \left[\Tr(\xi_\nu\Sigma \xi_\nu \Sigma^\dagger) 
         - \Tr(\xi_{\nu5}\Sigma \xi_{5\nu} \Sigma^\dagger) \right]
\nonumber \\ &&
+\, C_6 \sum_{\mu<\nu} 
\Tr(\xi_{\mu\nu}\Sigma \xi_{\nu\mu} \Sigma^\dagger) \,.
\label{eq:massterms}
\end{eqnarray}
The six unknown coefficients $C_i$ are all of size $\Lambda_{\rm TC}^6$.

In the continuum, the pions form a 15-plet of flavor $SU(4)$,
and are degenerate. On the lattice, states are classified by the
symmetries of the transfer matrix  and the Goldstone  pions
fall into 7 irreducible representations: four 3-dimensional representations
with flavors $\xi_i$, $\xi_{i5}$, $\xi_{ij}$ and $\xi_{i4}$,
and three 1-dimensional representations with flavors $\xi_4$, $\xi_{45}$ and $\xi_5$.

Close to both the chiral and continuum limits,
the pion masses are given by
\begin{equation}
M_\pi(T_a)^2 = 2 B m_q + a^2 \Delta(T_a) + O(a^2 m_q) + O(a^4) \,,
\label{eq:pimassform}
\end{equation}
with $\Delta(T_a) \sim \Lambda_{\rm TC}^4$ arising from $\CV_\chi^{(6)}$.
Since $\CV_\chi^{(6)}$ respects flavor $SO(4)$, the 15 Goldstone pions fall into $SO(4)$ representations:
\begin{eqnarray}
\hskip 0.3in \Delta(\xi_5) &=& 0 \,,
\label{eq:delta5}\\
\Delta(\xi_\mu) &=& {8 \over F^2} 
(C_1+C_2+C_3+3 C_4+C_5+3 C_6) \,, 
\label{eq:ximu}\\
\Delta(\xi_{\mu5}) &=&{8 \over F^2} 
(C_1+C_2+3 C_3+ C_4-C_5+3 C_6) \,,
\label{eq:ximu5}\\
\Delta(\xi_{\mu\nu}) &= &{8 \over F^2} (2C_3+2 C_4+4 C_6) \,.
\label{eq:ximunu}
\end{eqnarray}
 In the chiral limit
at finite lattice spacing the lattice irreducible representations  with flavors
$\xi_i$ and $\xi_4$ are degenerate,  those with flavors
$\xi_{i5}$ and $\xi_{45}$, and those with flavors
$\xi_{ij}$ and $\xi_{i4}$ are degenerate as well.
No predictions can be made for the ordering or splittings of the mass shifts.
We also cannot predict the {\em sign} of the shifts, although our
simulations indicate that they are all positive with the exponentially smeared staggered action
we use. This makes the existence of an Aoki phase~\cite{Lee:1999zxa} unlikely.

The method of ~\cite{Lee:1999zxa} has been generalized in a nontrivial way
to the $N_f > 4$ case~\cite{Aubin:2003mg,Aubin:2003uc} 
which we adopted in our calculations with
help from Bernard and Sharpe. The procedure cannot be reviewed here but it will be used in the interpretation
of our $N_f=8$ simulations.

\subsection{Finite volume analysis in the p-regime}

Three different regimes can be selected in simulations to identify the chirally broken phase from finite volume spectra and
correlators.
For a  lattice size $L_s^3\times L_t$ in euclidean space and in the limit $L_t \gg  L_s$, the conditions 
$F_{\pi}L_s > 1$ and 
$M_{\pi}L_s > 1$ select the the p-regime, in analogy 
with low momentum counting~\cite{Gasser:1987zq,Hansen:1990yg}. 

For arbitrary $N_f$, in the continuum and in infinite volume, the one-loop chiral corrections  to $M_\pi$ and $F_\pi$
of the degenerate Goldstone pions are given by
\begin{equation}
                        ~~~~  M^2_\pi = M^2  \biggl [1-\frac{M^2}{8\pi^2N_fF^2}ln\biggl (\frac{\Lambda_3}{M}\biggr ) \biggr ] ,
\label{eq:Mpi}                         
\end{equation}
\begin{equation}
                         F_\pi = F  \biggl [1+\frac{N_fM^2}{16\pi^2F^2}ln\biggl (\frac{\Lambda_4}{M}\biggr ) \biggr ] ,
\label{eq:Fpi}
\end{equation}
where $M^2=2B\cdot m_q$ and $F,B,\Lambda_3,\Lambda_4$ are four fundamental parameters of the chiral 
Lagrangian, and the small quark mass $m_q$ explicitly breaks the symmetry~\cite{Gasser:1983yg}. The chiral parameters
$F,B$ appear in the leading part of the Lagrangian in Eq.~(\ref{eq:Lagrange}), while $\Lambda_3,\Lambda_4$ 
enter in next order. There is the well-known GMOR relation
$\Sigma_{cond}=BF^2$ in the $m_q \rightarrow 0$ limit for the  chiral condensate per unit flavor~\cite{GellMann:1968rz}. 
It is important to 
note that the one-loop correction to the pion coupling constant $F_{\pi}$ is enhanced by a factor $N_f^2$ compared
to $M_{\pi}^2$.  The chiral expansion for large $N_f$ will break down for $F_{\pi}$ much faster for a given 
$M_{\pi}/F_{\pi}$ ratio.

The finite volume corrections to $M_\pi$ and $F_\pi$ are given in the p-regime
by
\begin{equation}
      M_\pi(L_s,\eta) = M_\pi  \biggl [1+\frac{1}{2N_f}\frac{M^2}{16\pi^2F^2}\cdot\tilde g_1(\lambda,\eta) \biggr ] ,
\label{eq:MpiL}
\end{equation}
\begin{equation}
     F_\pi (L_s,\eta)= F_\pi \biggl [1-\frac{N_f}{2}\frac{M^2}{16\pi^2F^2} \cdot\tilde g_1(\lambda,\eta) \biggr ] ,
\label{eq:FpiL}
\end{equation}
where $\tilde g_1(\lambda,\eta)$ describes the finite volume corrections with $\lambda=M\cdot L_s$ 
and aspect ratio $\eta=L_t/L_s$.
The form of $\tilde g_1(\lambda,\eta)$ is a complicated infinite sum which contains Bessel functions and requires numerical evaluation~\cite{Hansen:1990yg}. Eqs.~(\ref{eq:Mpi}-\ref{eq:FpiL}) provide the foundation of
the p-regime fits in our simulations.

\begin{figure}[ht!]
\begin{center}
\includegraphics[width=7cm,angle=0]{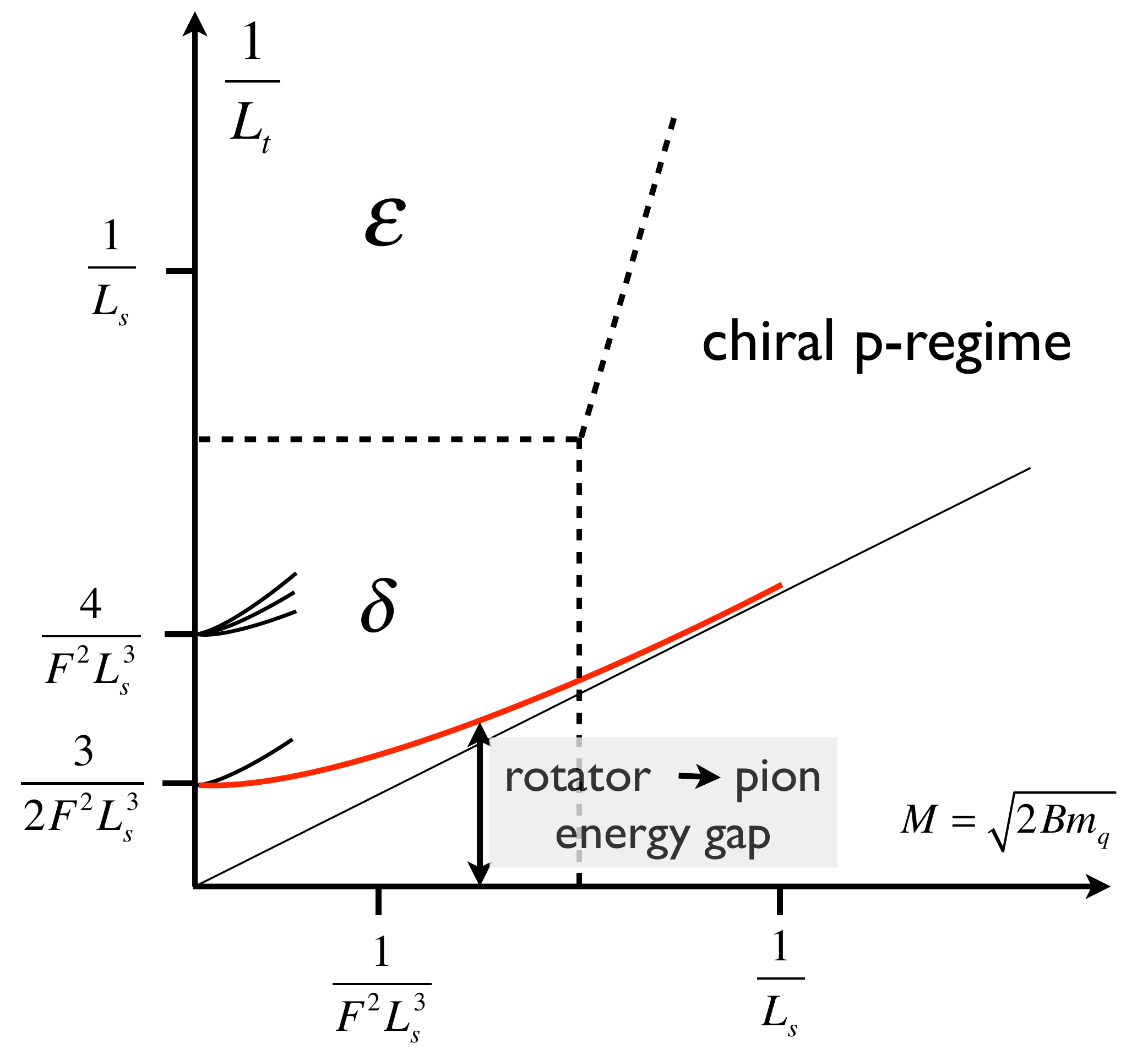} 
\end{center}
\caption{Schematic plot of the regions in which the three low energy chiral expansions are valid. 
The vertical axis shows the finite temperature scale (euclidean time in the path integral) which probes
the rotator dynamics of the $\delta$-regime and the $\epsilon$-regime. The first two low lying rotator levels are
also shown on the vertical axis for the simple case of  $N_f=2$. The fourfold degenerate lowest rotator excitation
at $m_q=0$
will split into an isotriplet state (lowest energy level), which evolves into the p-regime pion as $m_q$ increases, and into an isosinglet
state representing a multi-pion state in the p-regime. Higher rotator excitations have similar interpretations.}   
\label{fig:landscape}
\end{figure}

\subsection{$\delta$-regime and $\epsilon$-regime}

At fixed $L_s$ and in cylindrical geometry $L_t/L_s \gg 1$,  a crossover occurs from the p-regime 
to the $\delta$-regime when $m_q \rightarrow 0$, as shown in Fig.~\ref{fig:landscape}. 
The dynamics is dominated by the rotator states of the chiral condensate
in this limit~\cite{Leutwyler:1987ak} which is characterized by the conditions $FL_s > 1$ and $ML_s \ll 1$. 
The densely spaced rotator spectrum scales with 
gaps of  the order $\sim 1/F^2L_s^3$, and at $m_q=0$ the chiral symmetry is apparently  restored. However, the
rotator spectrum, even at $m_q=0$ in the finite volume, will  signal that the infinite system is 
in the chirally broken phase for the particular parameter set of the Lagrangian. This is often misunderstood 
in the interpretation of lattice simulations.
Measuring finite energy levels
with pion quantum numbers at fixed $L_s$ in the  $m_q \rightarrow 0$ limit is not a 
signal for chiral symmetry restoration of the infinite
system~\cite{Deuzeman:2009mh}.
\begin{figure}[h!]
\begin{center}
\includegraphics[width=8cm,angle=0]{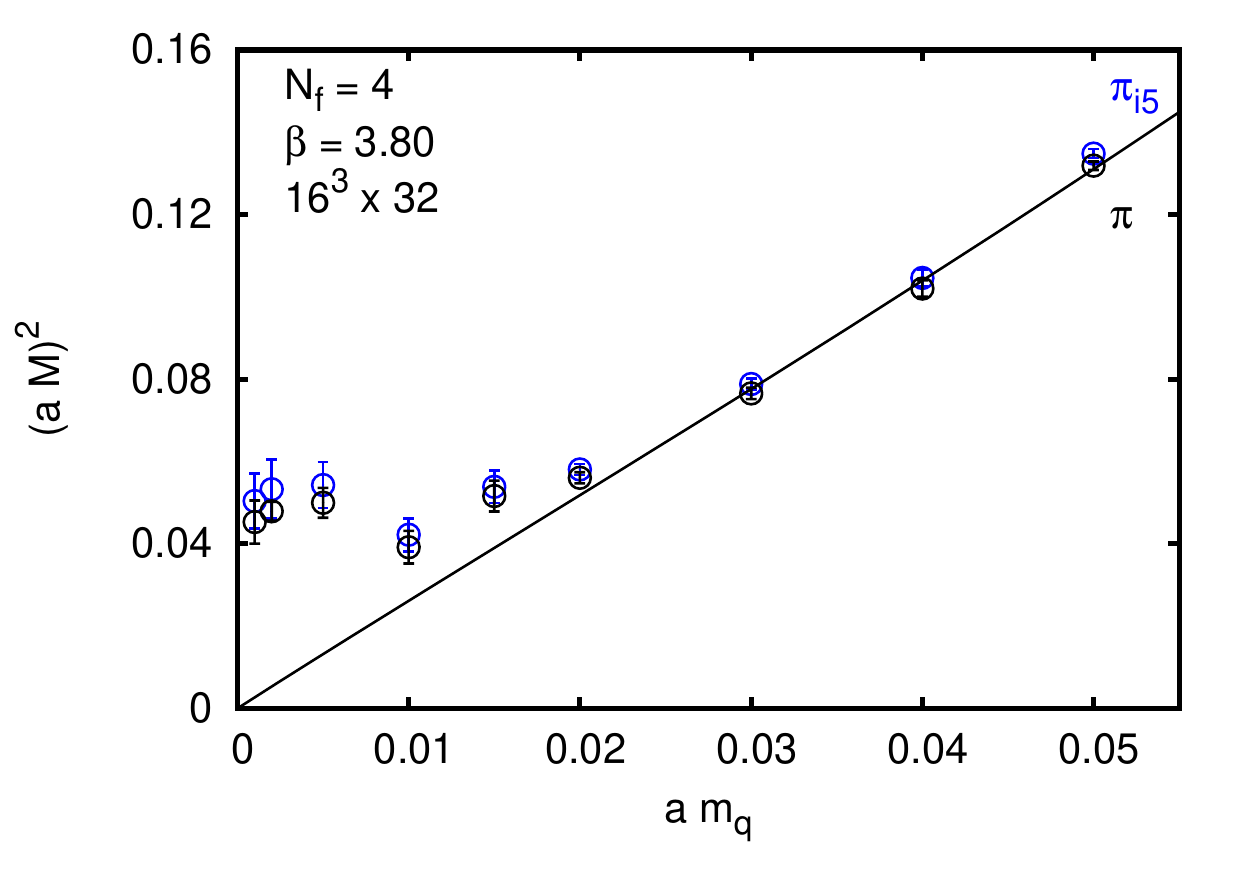}
\end{center}
\caption{The crossover from the p-regime to the $\delta$-regime is shown for the $\pi$ and $\pi_{i5}$
states at $N_f=4$. }   
\label{fig:delta}
\vskip -0.1in
\end{figure}
\begin{figure*}[ht!]
\begin{center}
\begin{tabular}{ccc}
\includegraphics[height=4.0cm]{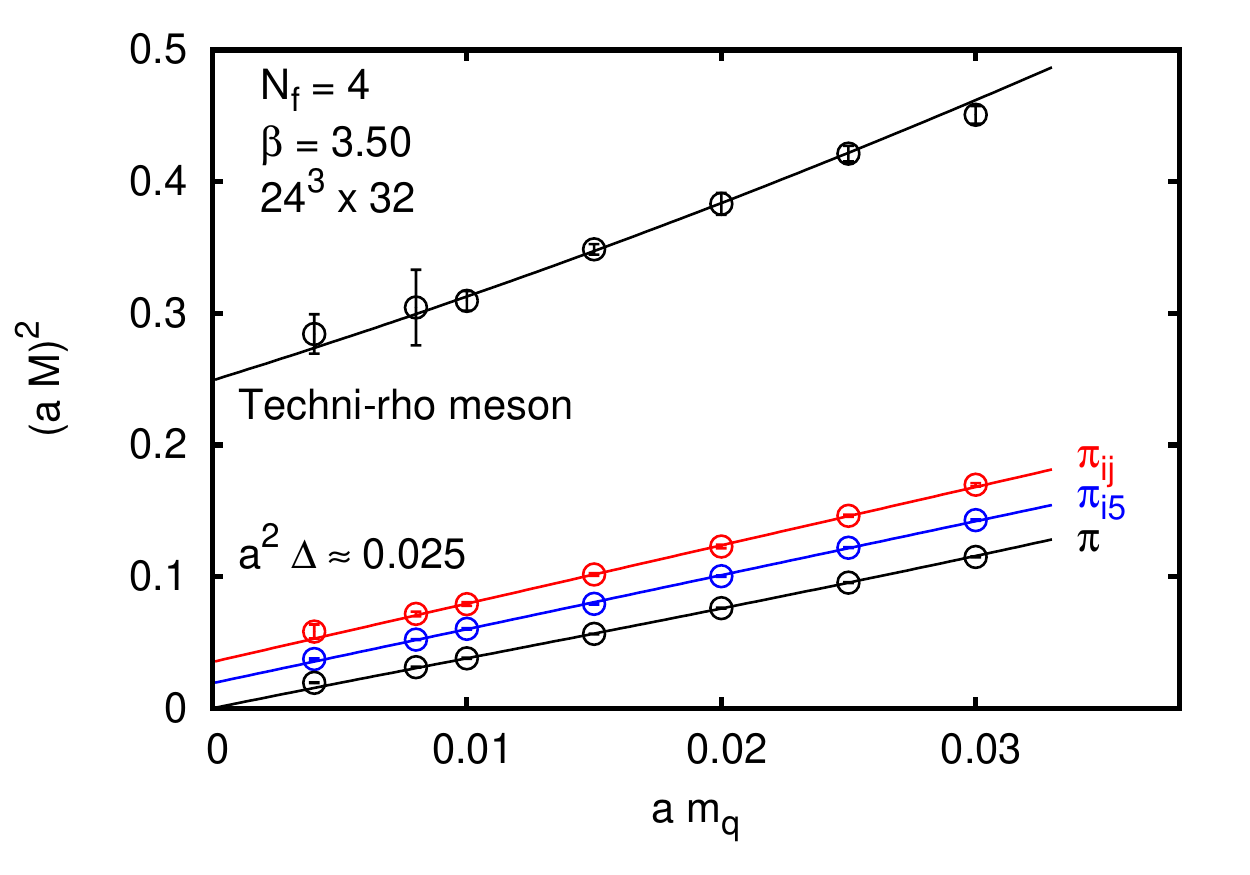}&
\includegraphics[height=4.0cm]{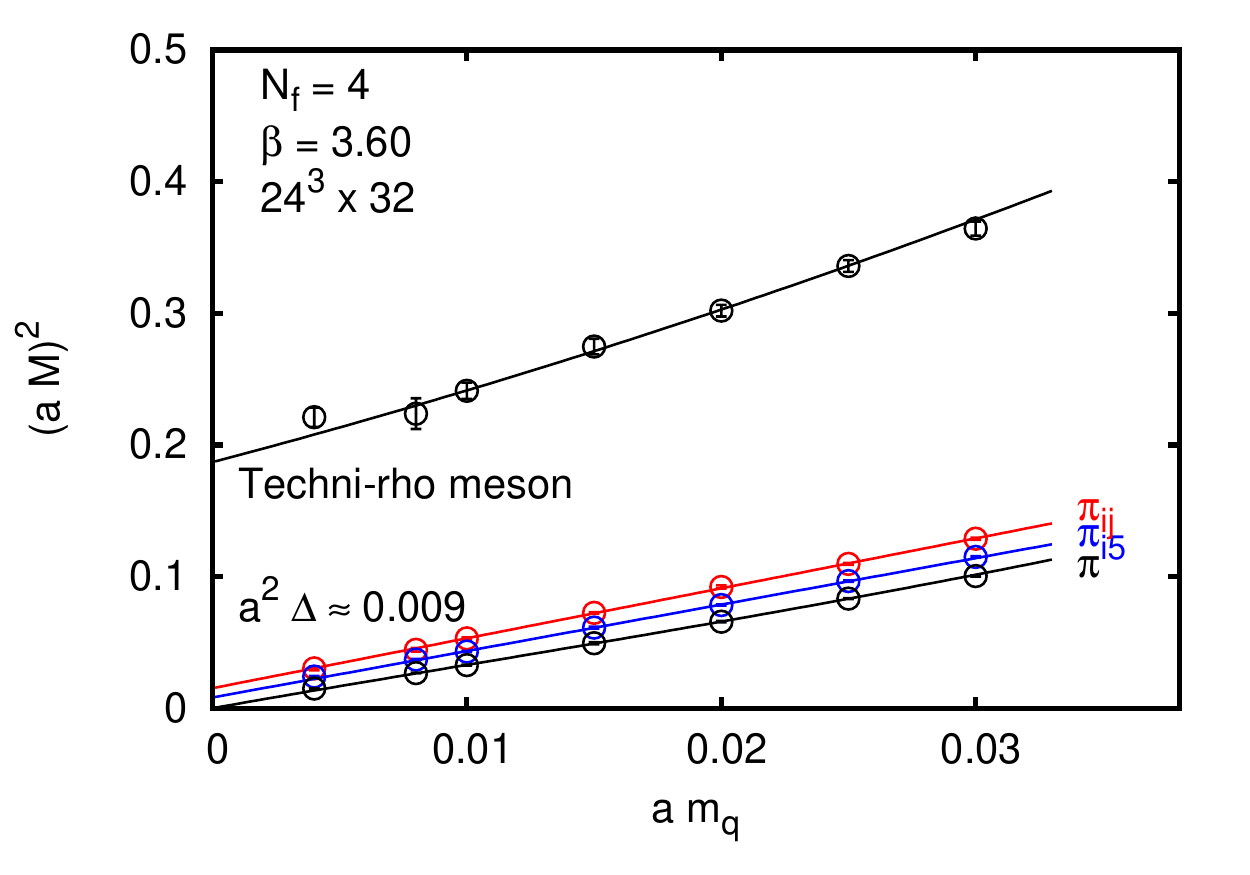}&
\includegraphics[height=4.0cm]{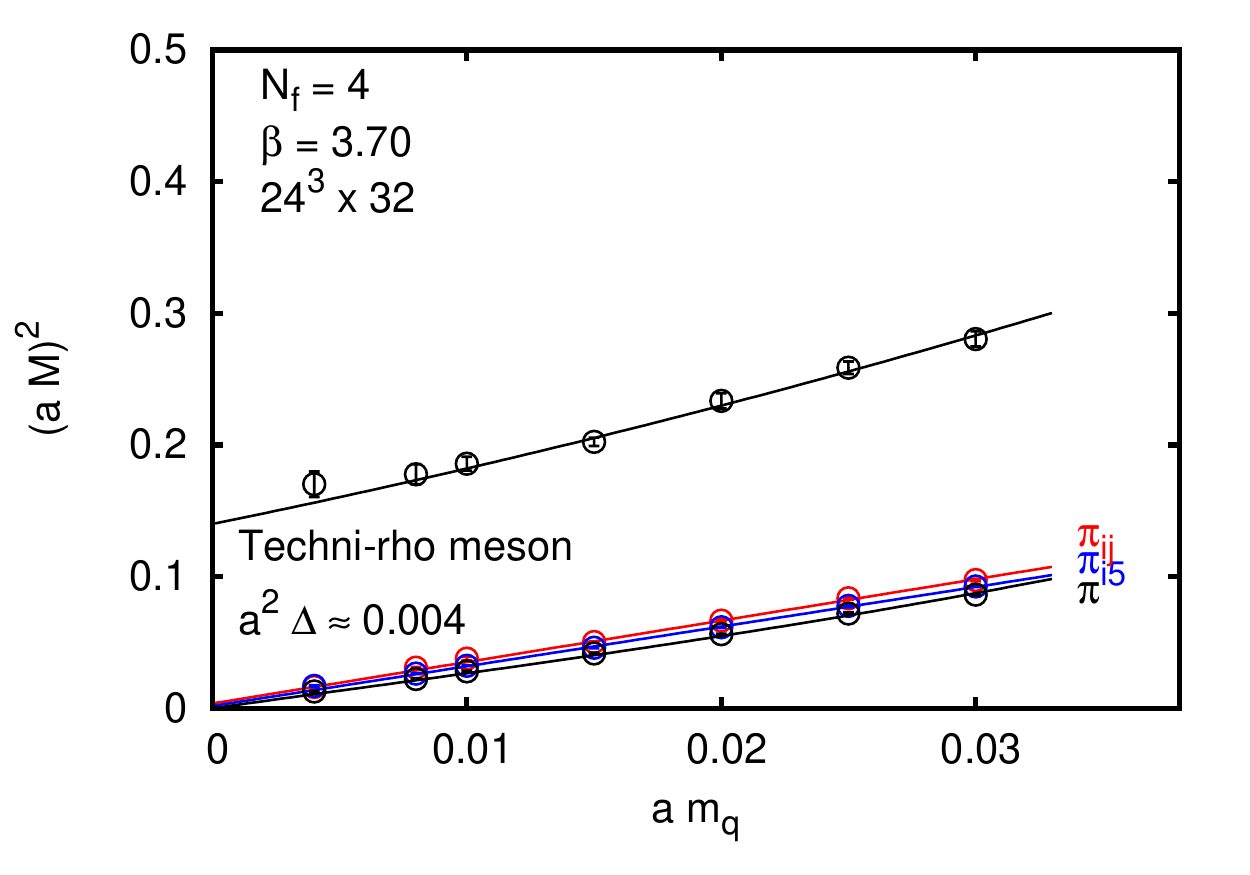}\\
\includegraphics[height=4.0cm]{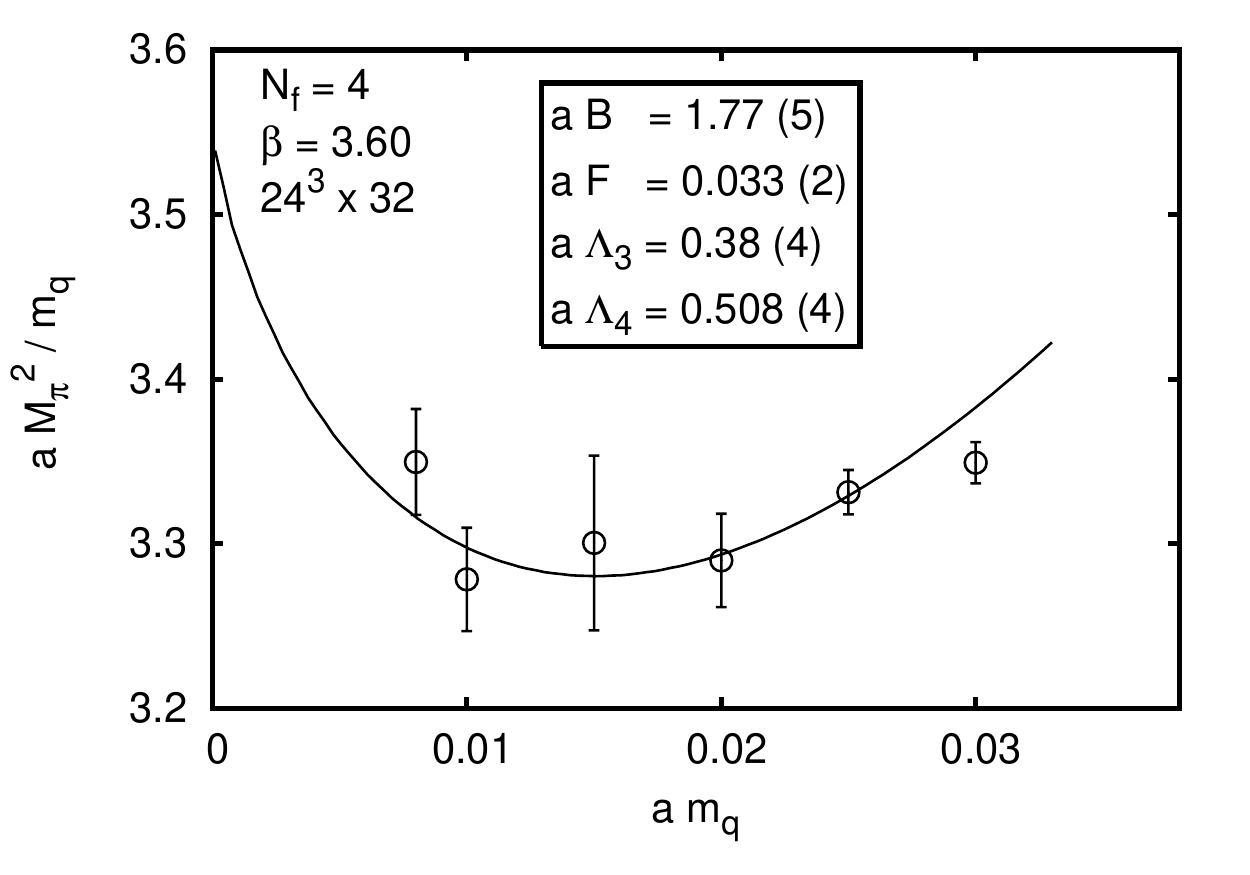}
&\includegraphics[height=4.0cm]{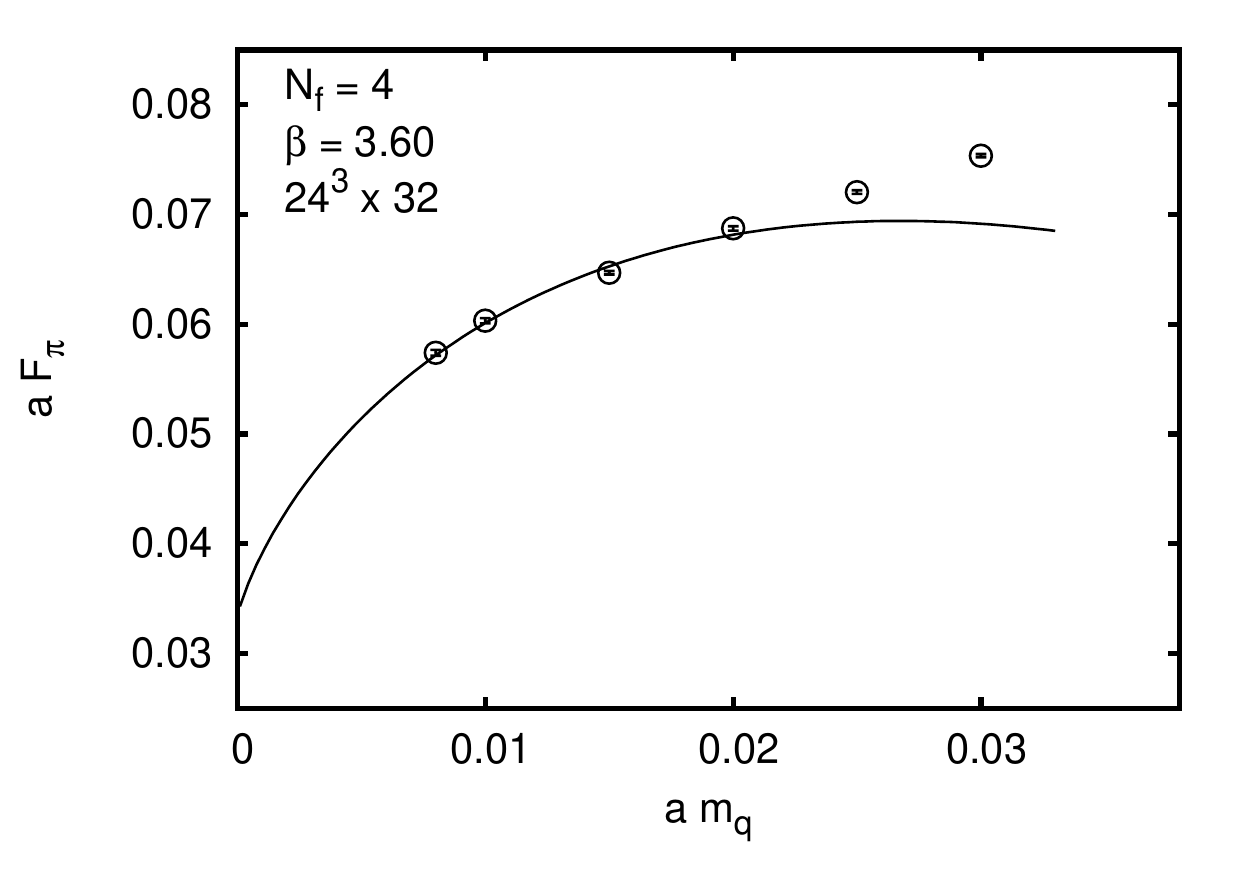}
&\includegraphics[height=4.0cm]{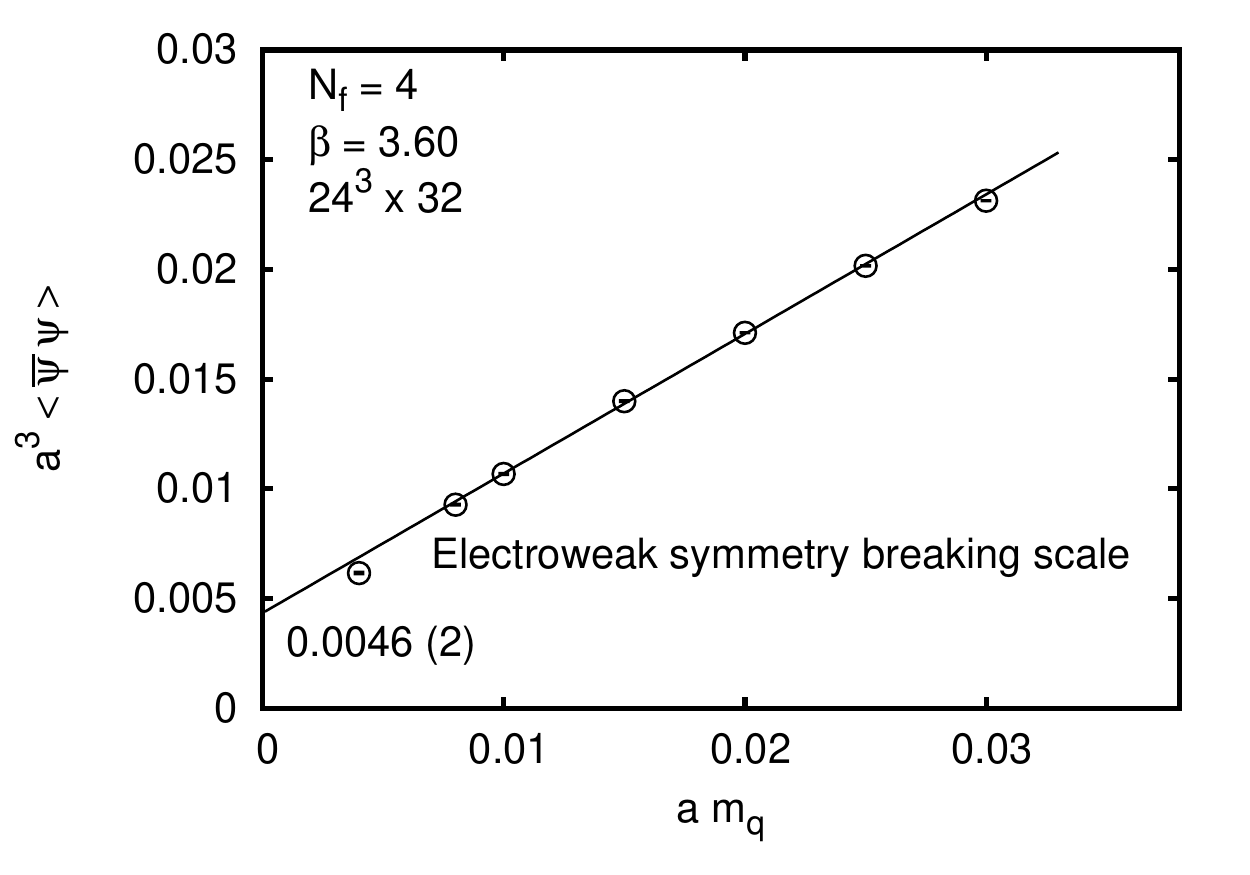}\\
\includegraphics[height=4.0cm]{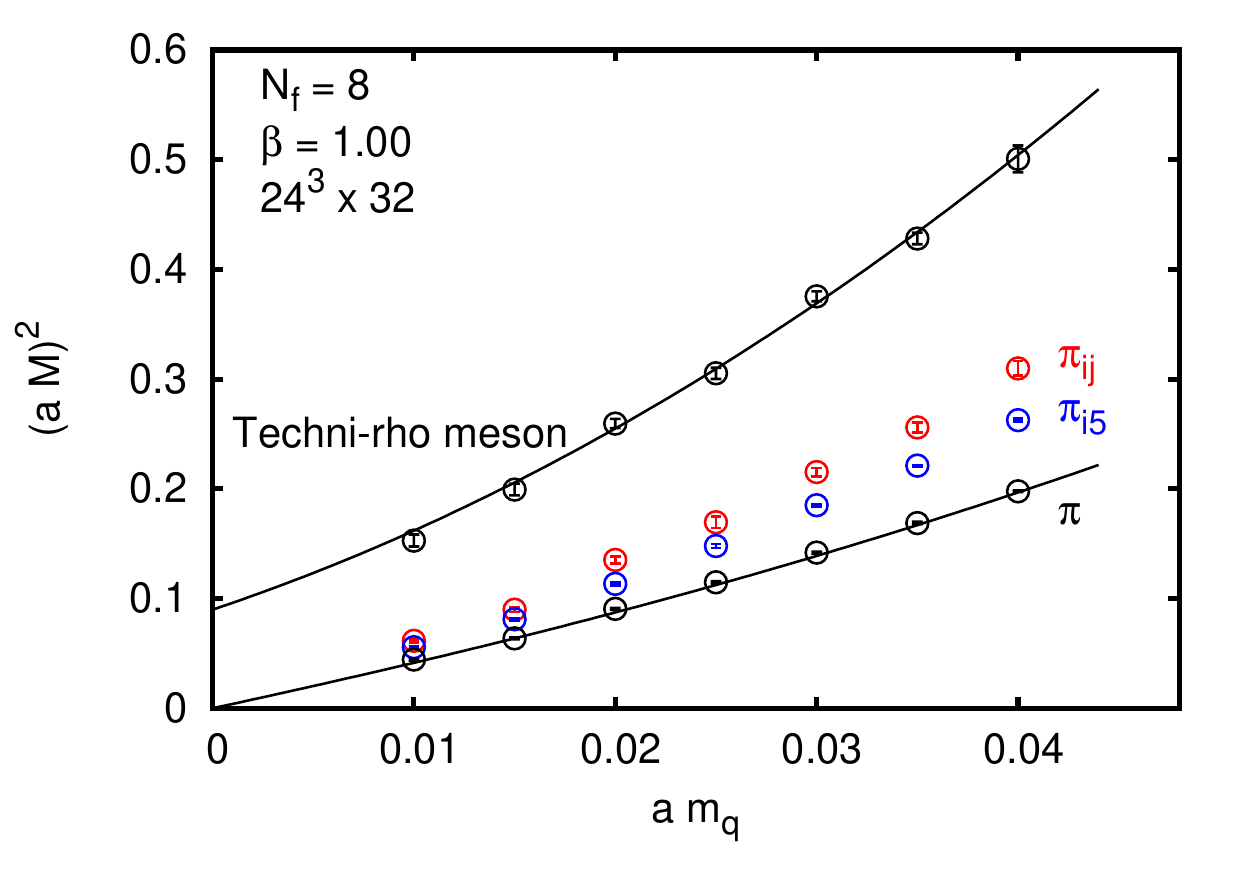}&
\includegraphics[height=4.0cm]{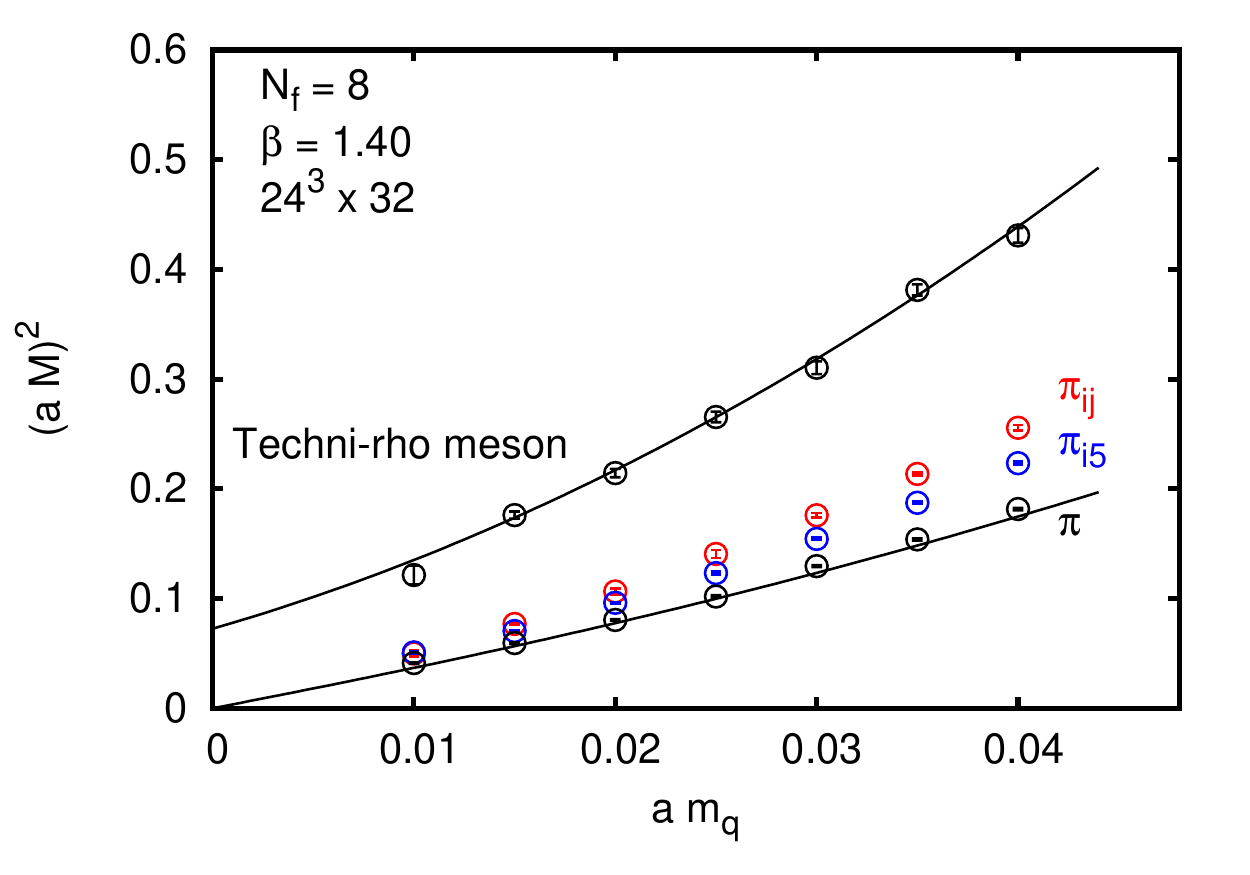}&
\includegraphics[height=4.0cm]{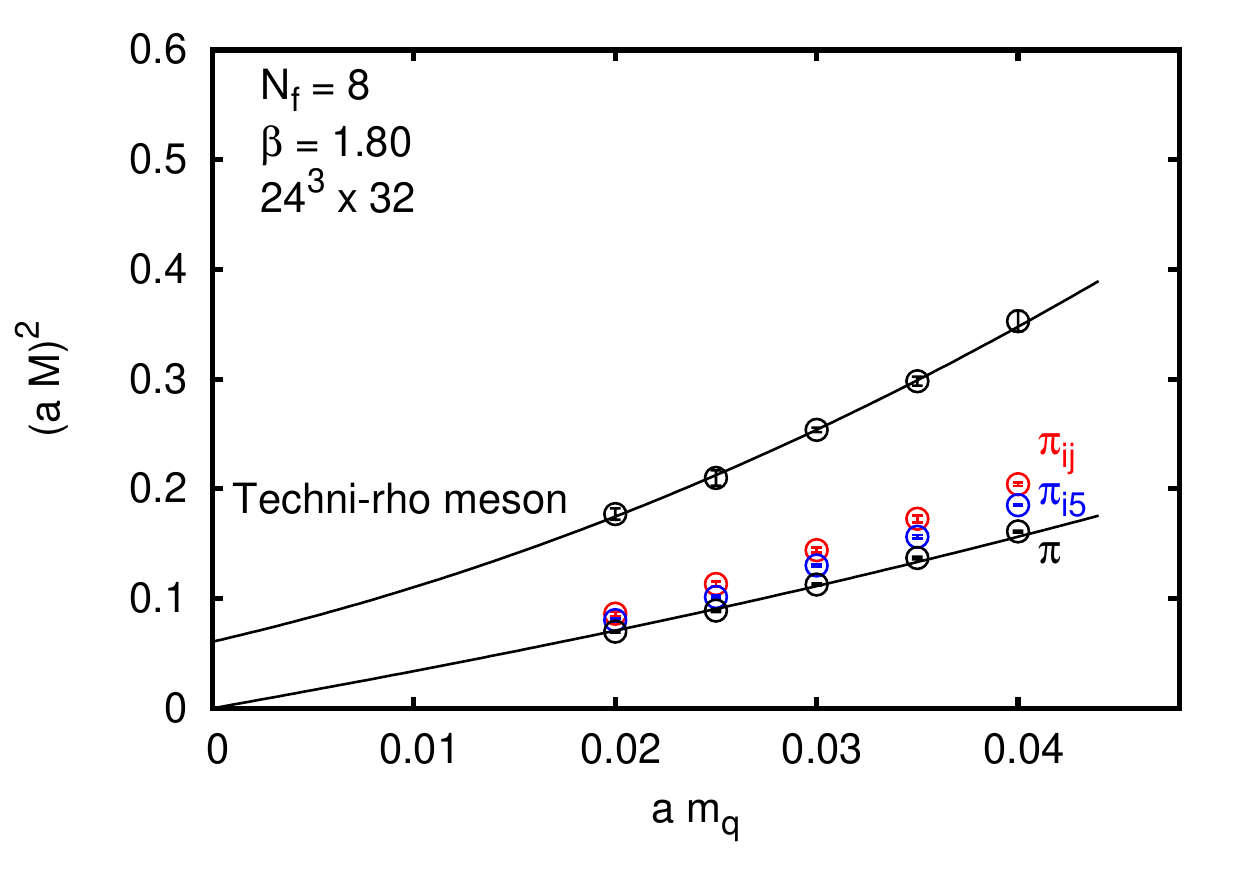}\\
\includegraphics[height=4.0cm]{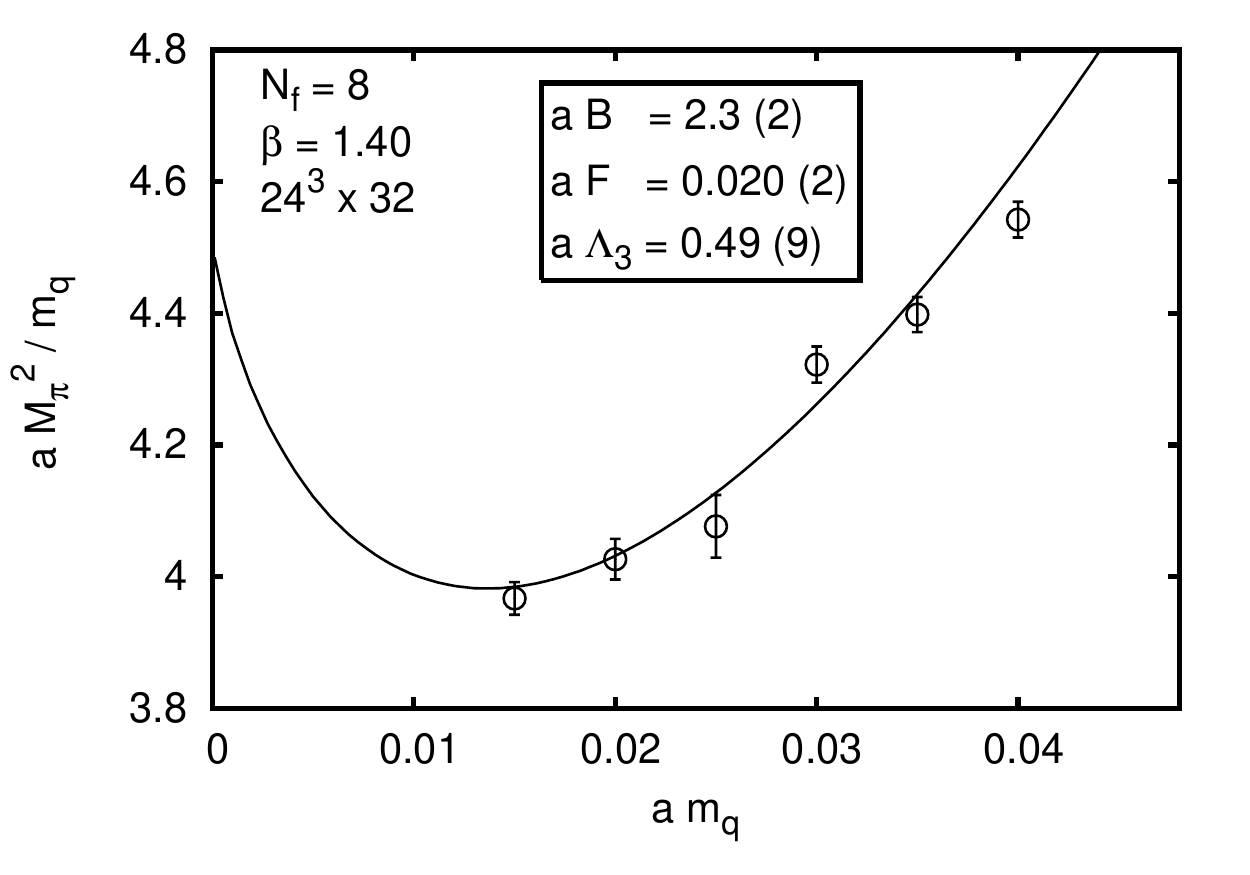}
&\includegraphics[height=4.0cm]{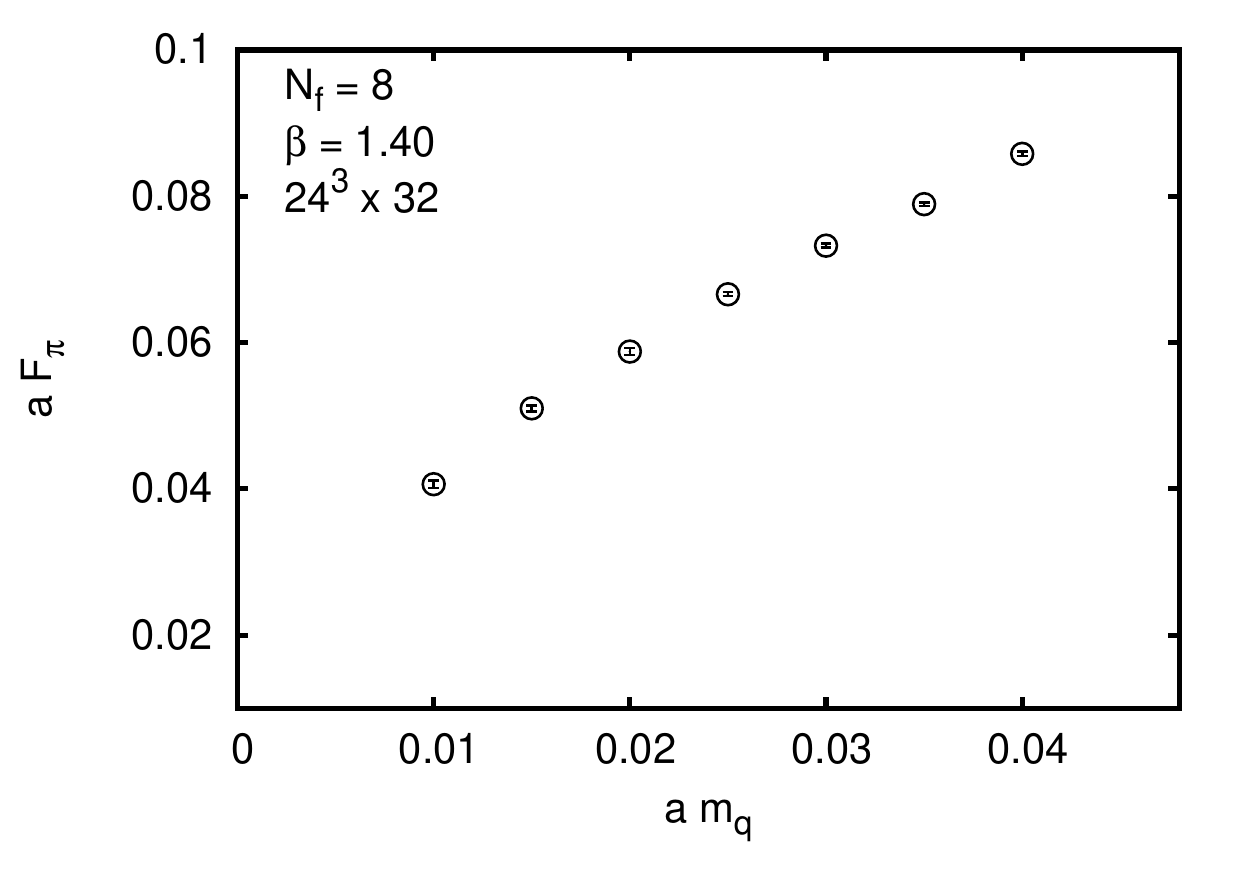}
&\includegraphics[height=4.0cm]{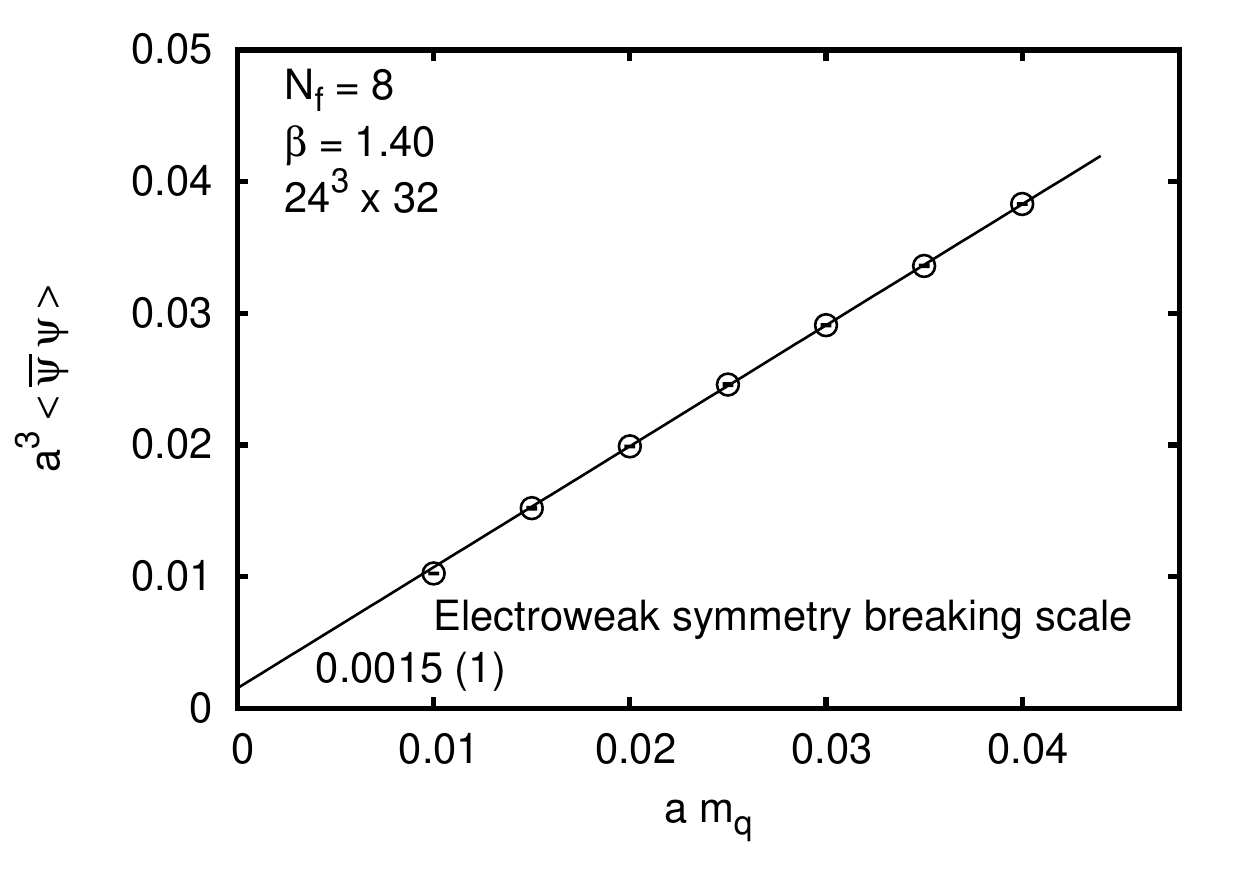}\\
\includegraphics[height=4.0cm]{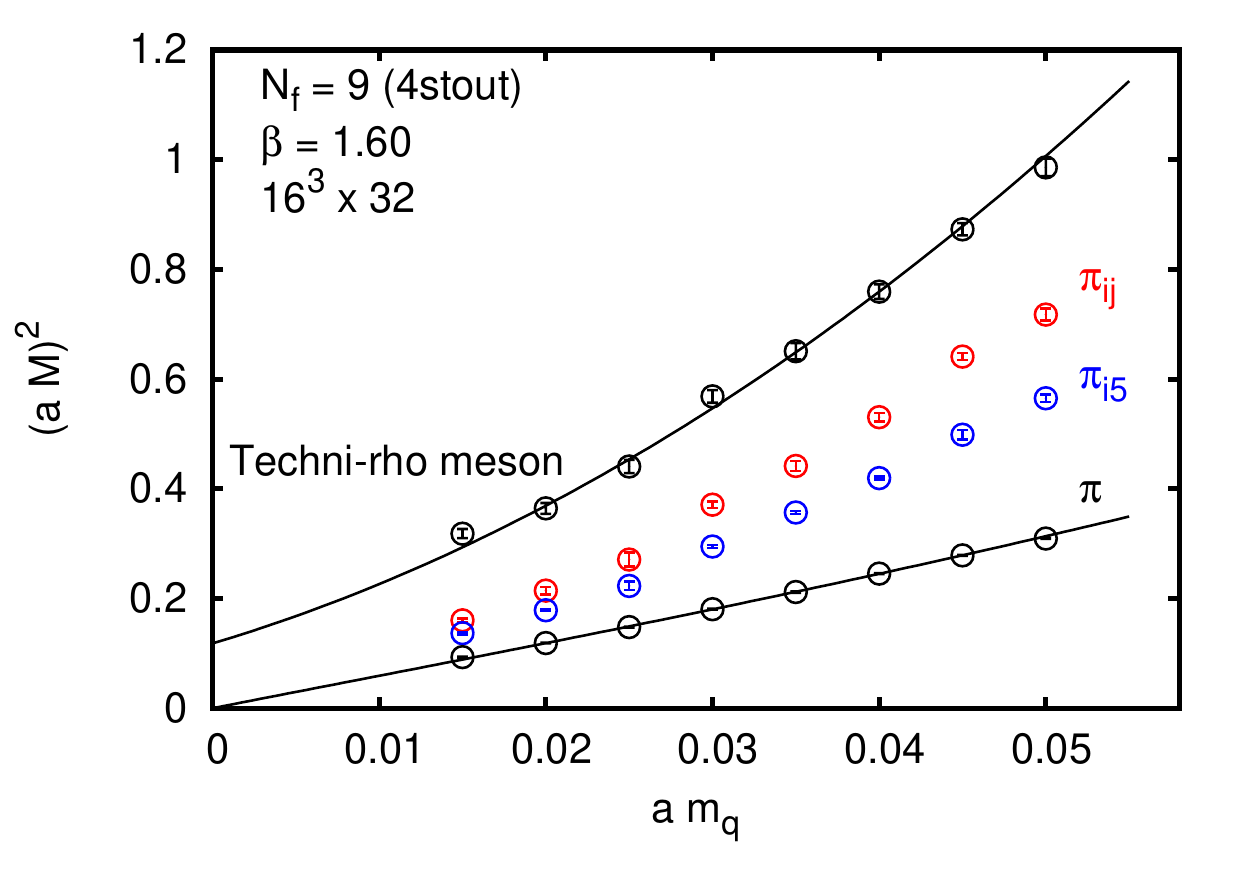}
&\includegraphics[height=4.0cm]{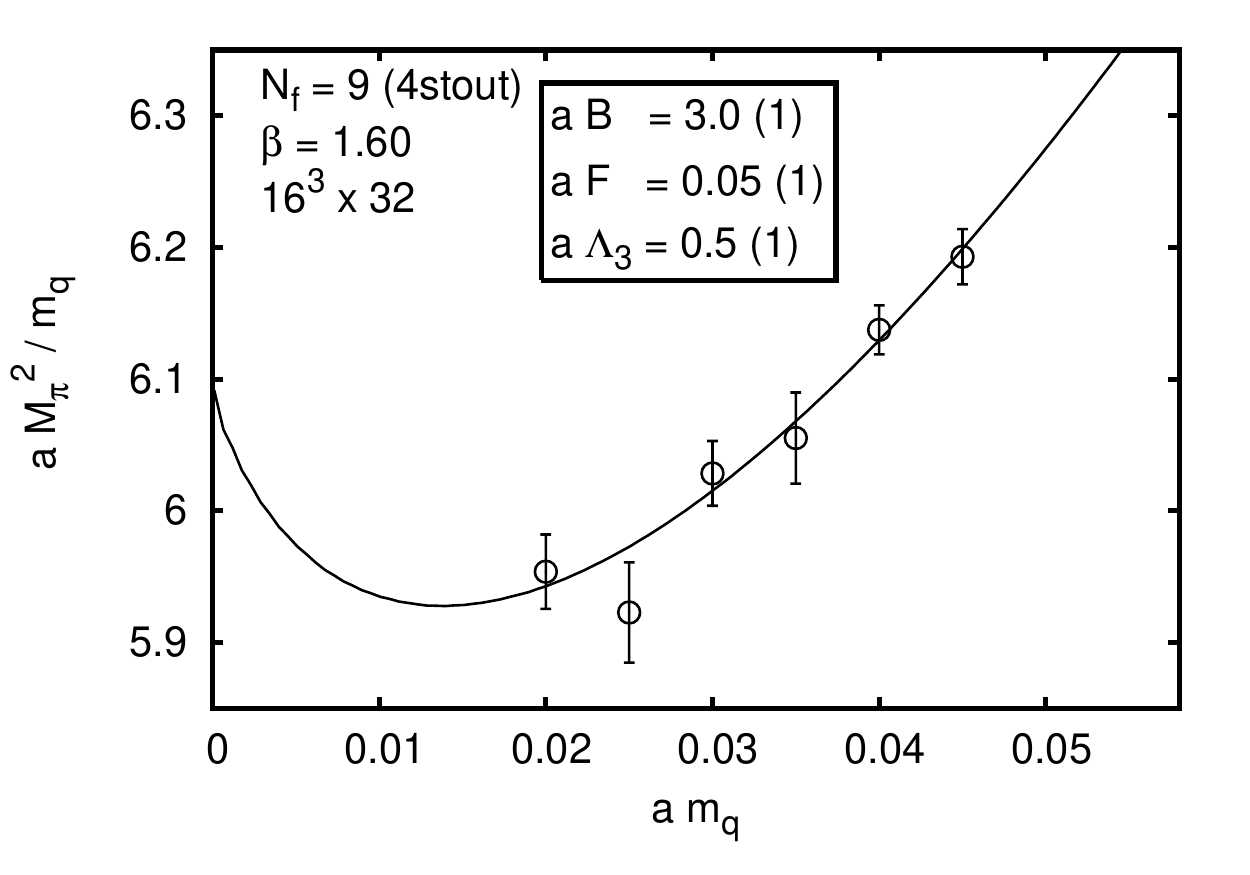}
&\includegraphics[height=4.0cm]{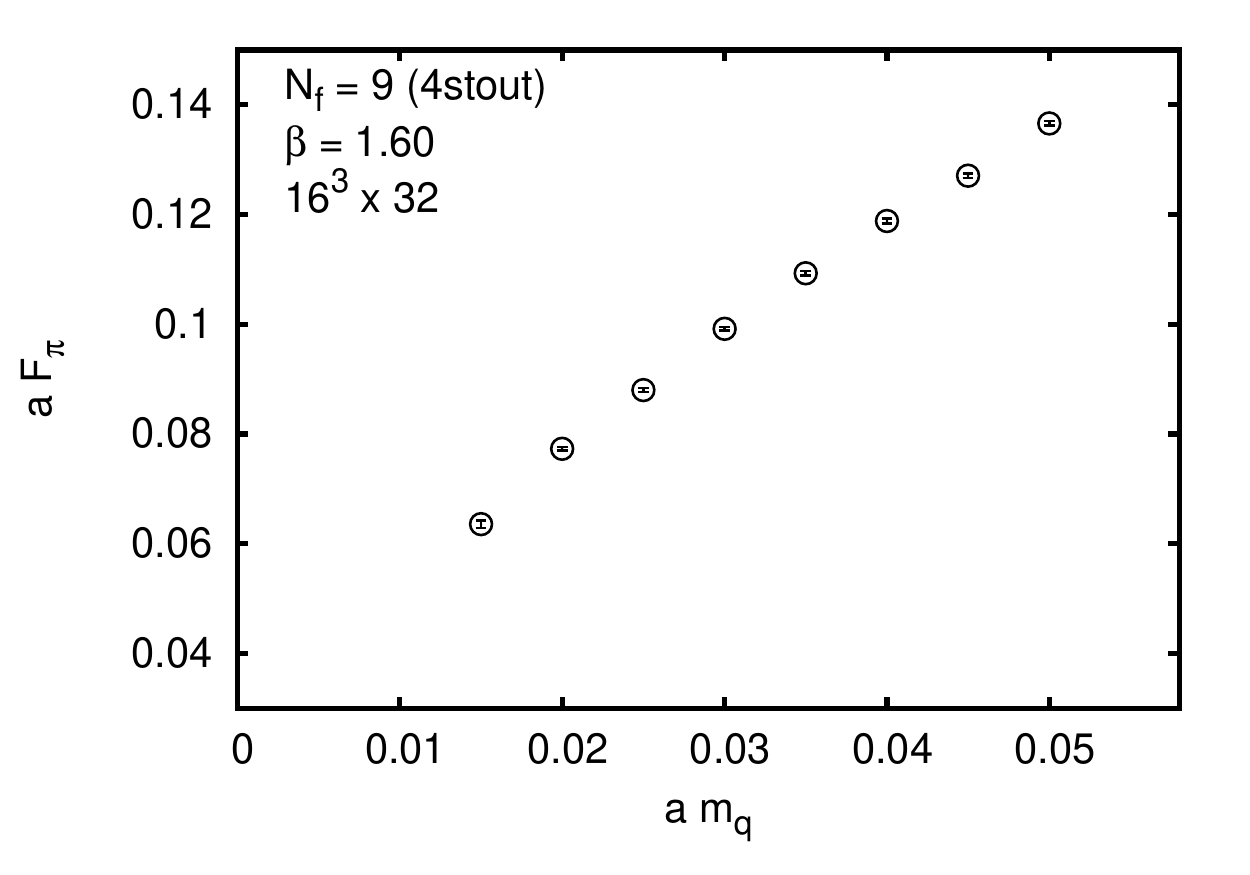}
\end{tabular}
\end{center}
\caption{The first two rows of the composite figure show $N_f=4$ simulation results in the p-regime. The first row 
depicts the collapsing pion spectrum and the techni-rho as the continuum limit is approached. The second row shows
the chiral fits to $M^2_\pi/m_q$ and $F_\pi$ based on Eqs.~(\ref{eq:Mpi}-\ref{eq:FpiL}).  The range
$m_q=0.008-0.025$ is used in the fitting procedure. The approximately linear
behavior of the chiral condensate $\langle \bar\psi\psi\rangle$ is also shown in the second row. The third and fourth rows 
summarize the simulation results for $N_f=8$. The third row shows  the collapsing pion spectrum and the techni-rho as 
the continuum limit is approached. The chiral fit to $M^2_\pi/m_q$ is shown based on Eq.~(\ref{eq:Mpi}) only since 
the $F_\pi$ data points are outside the convergence range of the chiral expansion. 
The range
$m_q=0.015-0.035$ is used in the fitting procedure.
The fifth row illustrates our first simulation results for $N_f=9$. It shows the split
pion spectrum, chiral fit to $M^2_\pi/m_q$ and the  $F_\pi$ data points are outside the convergence range of the chiral 
expansion. The range
$m_q=0.02-0.04$ is used in the fitting procedure.}   
\label{fig:Nf4}
\vskip -0.1in
\end{figure*}

If $L_t \sim L_s$ under the conditions $FL_s > 1$ and $ML_s \ll 1$, the system will be driven into
the $\epsilon$-regime which can be viewed as the high temperature limit of 
the $\delta$-regime quantum rotator.  Although the $\delta$-regime and $\epsilon$-regime have
an overlapping region, there is an important difference in their dynamics. In the $\delta$-regime of the quantum rotator, the 
zero spatial momentum of the pion field $U(x)$ dominates with  time-dependent quantum dynamics. 
The $\epsilon$-regime is dominated by
the four-dimensional zero momentum  mode of the  chiral Lagrangian. 

We report simulation results of all three regimes in the chirally broken phase of the technicolor models we investigate. The 
analysis of the three regimes
complement each other and provide cross-checks for the correct identification of the phases. First, we will probe 
 Eqs.(\ref{eq:Mpi}-\ref{eq:FpiL}) in the p-regime, and follow with the study of Dirac spectra and RMT eigenvalue distributions in the
 $\epsilon$-regime. The spectrum in the $\delta$-regime is used as a signal to monitor p-regime spectra as $m_q$
 decreases.  Fig.~\ref{fig:delta} is an  illustrative example for this crossover in our simulations.

\section{Simulations results in the p-regime}

The tree level improved Symanzik  gauge action was used in our simulations. The link variables in the staggered fermion matrix
were exponentially smeared with  six stout steps at $N_f=4,8$ and four stout steps at $N_f=9$. The RHMC algorithm was deployed
in all runs but rooting of the fermion determinant only affected the $N_f=9$ simulations. 
The results shown in Fig.~\ref{fig:Nf4} are from the p-regime of the chirally broken phase with the 
conditions  $M_\pi\cdot L_s \gg 1$ and $F_\pi\cdot L \sim 1$  when the chiral condensate begins to follow
the expected behavior of infinite volume chiral perturbation theory from Eqs.~(\ref{eq:Mpi},\ref{eq:Fpi})
with calculable finite volume corrections from Eqs.~(\ref{eq:MpiL},\ref{eq:FpiL}). 

The $N_f=4$ simulations work in the p-regime as expected. 
The pion spectrum is clearly separated from the technicolor scale of the
$\rho$-meson whose quadratic fit is just to guide the eye.
Moving towards the continuum limit with increasing $\beta=6/g^2$, we see the split pion spectrum collapsing onto
the true Goldstone pion. The true Goldstone pion 
and two additional split pion states are shown.
$\Delta$ is the measure of the  small quadratic pion mass splittings in lattice units. Their origin was discussed in Section 2
in Eqs.~(\ref{eq:delta5}-\ref{eq:ximunu}). The spectrum is parallel and the gaps appear to be equally spaced consistent with
the earlier observation in QCD where the $C_4$ term seems to dominate taste breaking 
accounting for the equally spaced pion levels~\cite{Lee:1999zxa}.
The simultaneous chiral fit of  $M^2_\pi/m_q$ and $F_\pi$ based on Eqs.~(\ref{eq:Mpi}-\ref{eq:FpiL}) works
when the chiral loop term corrects the tree level value of $M^2_\pi/m_q=2B$.
This is a chirally broken phase and the picture holds in the $m_q\rightarrow 0$ limit.
The fit to determine the $N_f=4$ chiral condensate for $m_q=0$  is shown in the second row on the right. It sets the
scale of electroweak symmetry breaking in the Higgs mechanism.

As we move to the $N_f=8$ p-regime simulations summarized in the third and forth rows 
of Fig.~\ref{fig:Nf4} we observe the weakening of the chiral condensate and 
increased difficulties in passing the chiral tests. 
The pion spectrum is still clearly separated from the technicolor scale of the
$\rho$-meson.
Moving towards the continuum limit with increasing $\beta=6/g^2$, we see the split pion spectrum collapsing toward
the true Goldstone pion with a new distinguished feature. The true Goldstone pion 
and two additional split pion states are shown with different slopes as $m_q$ increases.
Towards $m_q=0$ the pion spectrum is collapsed at fixed gauge coupling, indicating that the effects 
of leading order taste breaking operators, 
the generalization
of those from $N_f=4$ to $N_f=8$ as discussed in Section 2,  are
smaller than at $N_f=4$ in the explored coupling constant range. 
This is somewhat unexpected and unexplained. Next to leading order 
taste breaking operators are responsible for the spread of the slopes and they seem to dominate. 
They were identified in Eq.~(\ref{eq:pimassform}) as the last two terms.
It is reassuring to
see that this structure is  collapsing as we move toward the continuum limit. We analyzed this pattern within staggered
perturbation theory in its generalized form beyond four flavors~\cite{Aubin:2003mg,Aubin:2003uc}. 
The simultaneous chiral fit of  $M^2_\pi/m_q$ and $F_\pi$ based on Eqs.~(\ref{eq:Mpi}-\ref{eq:FpiL}) cannot be
done at $N_f=8$ within the reach of the largest lattice sizes we deploy since the value of $aF$ is too small even at L=24
for coupling constants where taste breaking drops to an acceptable level. 
The chiral fit to $M^2_\pi/m_q$ is shown based on Eq.~(\ref{eq:Mpi}) only since 
the $F_\pi$ data points are outside the convergence range of the chiral expansion. We would need much bigger lattices to
drop further down in the p-regime with $m_q$ to the region where the simultaneous fit could be made.
It is also important to note that the chiral
condensate is very small in the $m_q\rightarrow 0$ limit in the region where taste breaking is not large. This is shown in
row four of  Fig.~\ref{fig:Nf4} on the right side.

The $N_f=9$ p-regime simulations are summarized in the fifth row 
of Fig.~\ref{fig:Nf4} where we observe the continued weakening of the chiral condensate and the
increased difficulties in passing the chiral tests. 
The pion spectrum is still clearly separated from the technicolor scale of the
$\rho$-meson. Moving towards the continuum limit to see the split pion spectrum collapsing toward
the true Goldstone pion is increasingly difficult. The true Goldstone pion 
and two additional split pion states are shown again with different slopes as $m_q$ increases. Forcing
the collapse of the split pion spectrum will require larger lattices with smaller gauge couplings.
The trends and the underlying explanation is very similar to the $N_f=8$ case. 
The chiral fit to $M^2_\pi/m_q$ is shown based on Eq.~(\ref{eq:Mpi}) only since 
the $F_\pi$ data points are outside the convergence range of the chiral expansion.

In summary, we have shown that  according to p-regime tests the $N_f=4,8,9$ systems are all in the chirally 
broken phase close to the continuum limit.
Currently we are investigating $N_f=9,10,11,12$ on larger lattices to determine the lower edge of the conformal window.
Lessons from the Dirac spectra and RMT to complement p-regime tests are discussed in the next section including comments 
about the controversial $N_f=12$ case.
\begin{figure*}[ht]
\begin{center}
\begin{tabular}{ccc}
\includegraphics[height=4.cm]{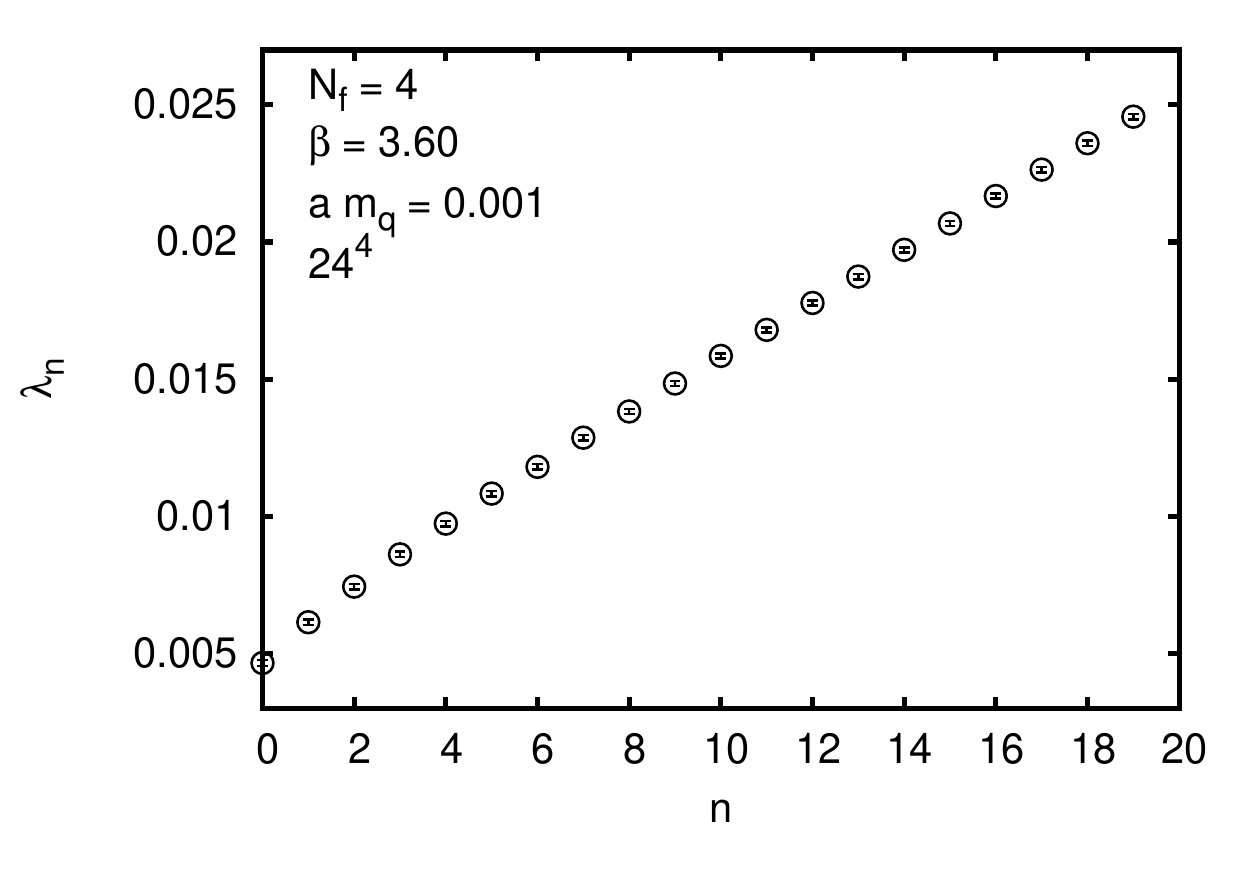}&
\includegraphics[height=4.cm]{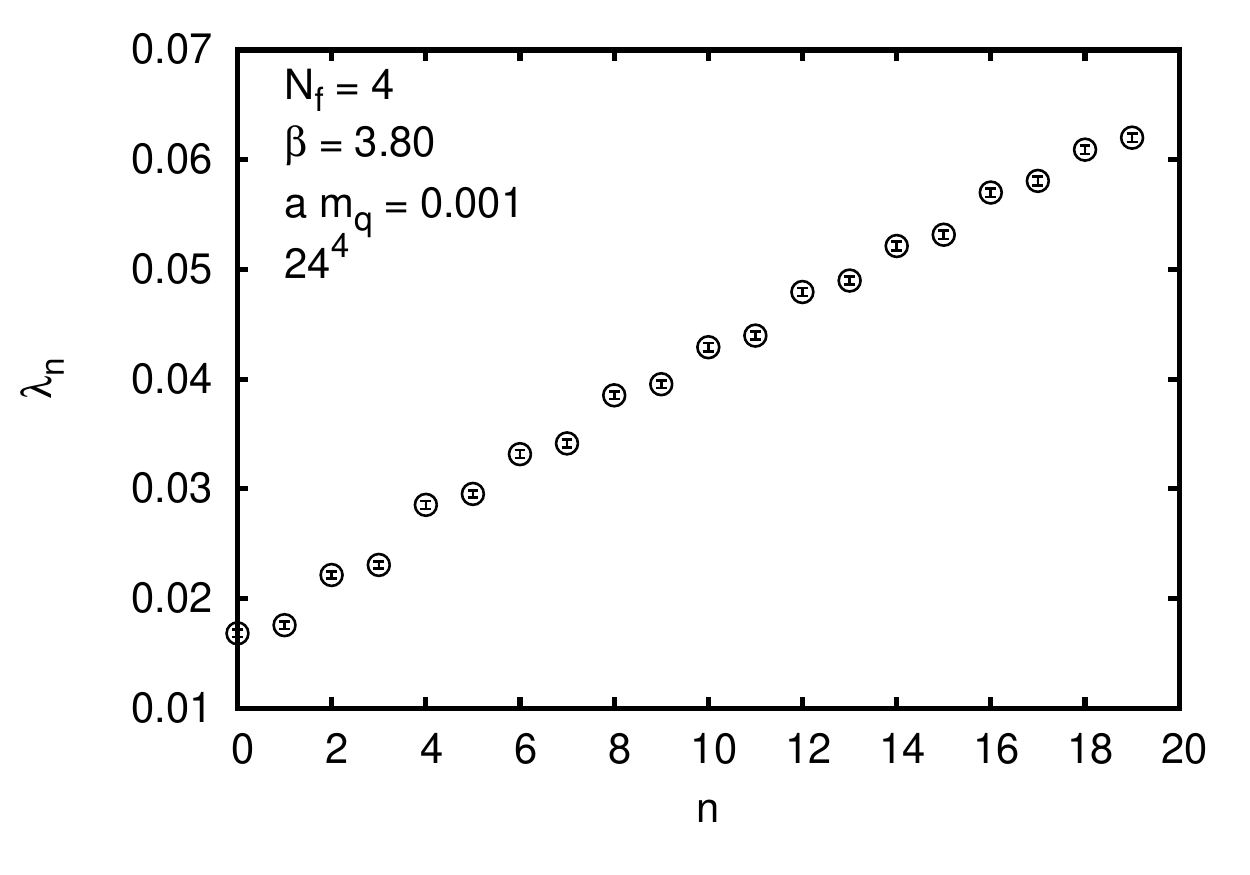}&
\includegraphics[height=4.cm]{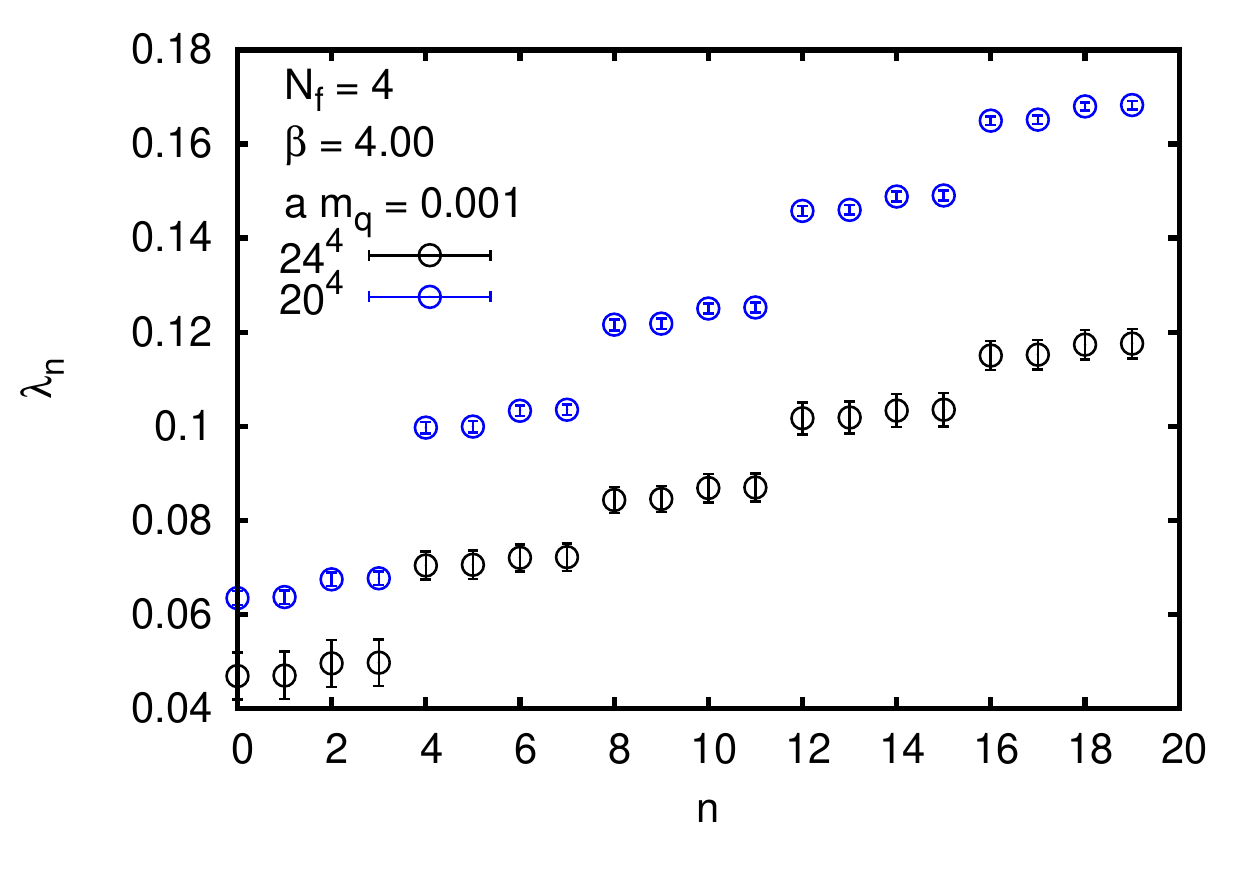}\\
\includegraphics[height=4.cm]{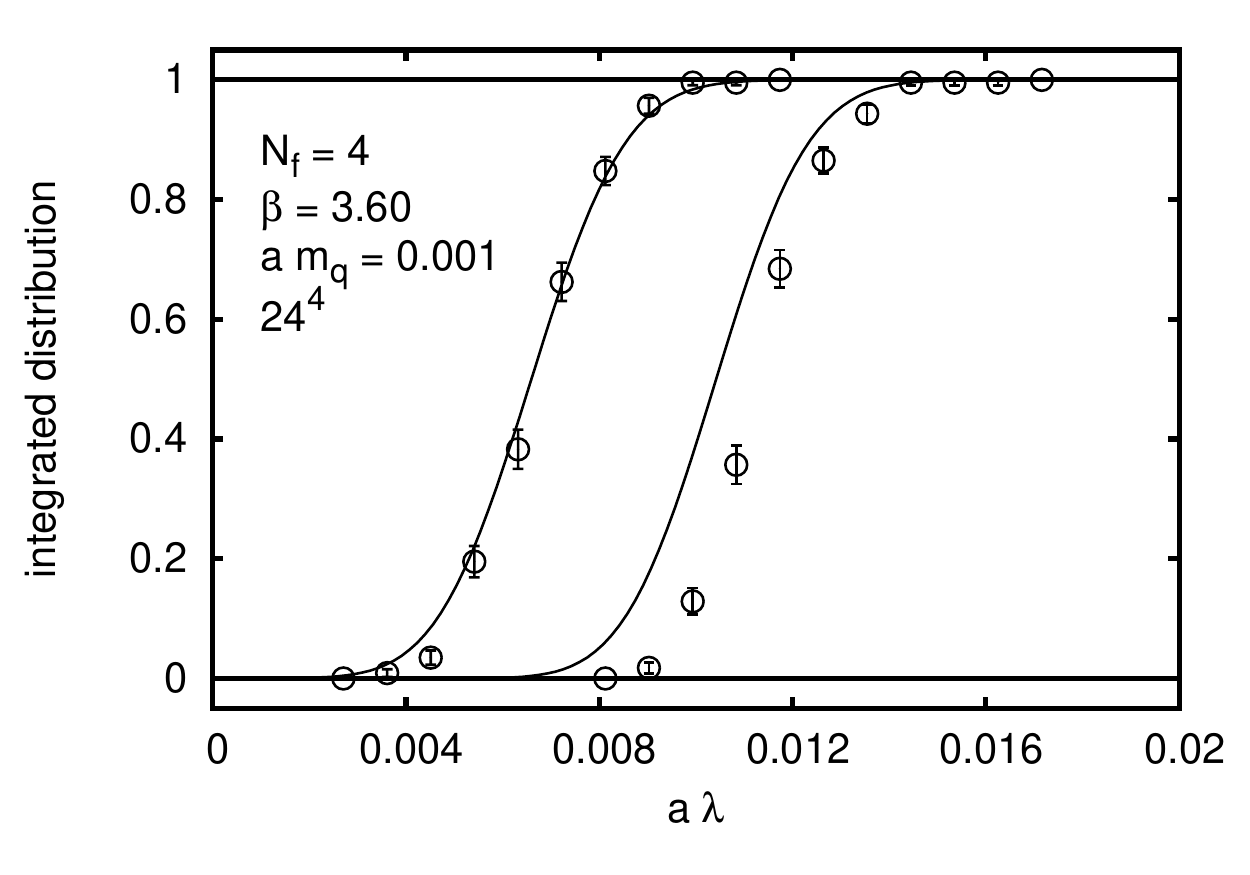}
&\includegraphics[height=4.cm]{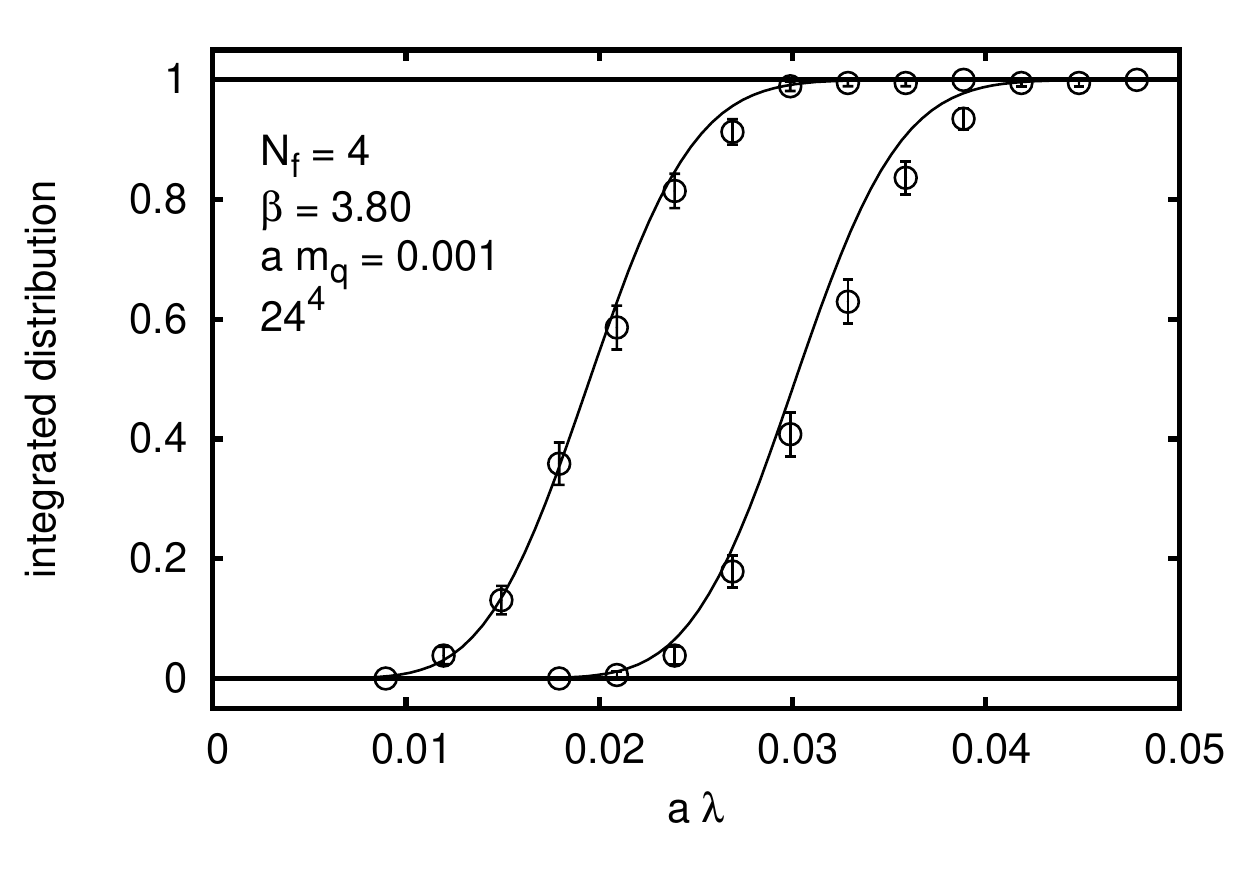}
&\includegraphics[height=4.cm]{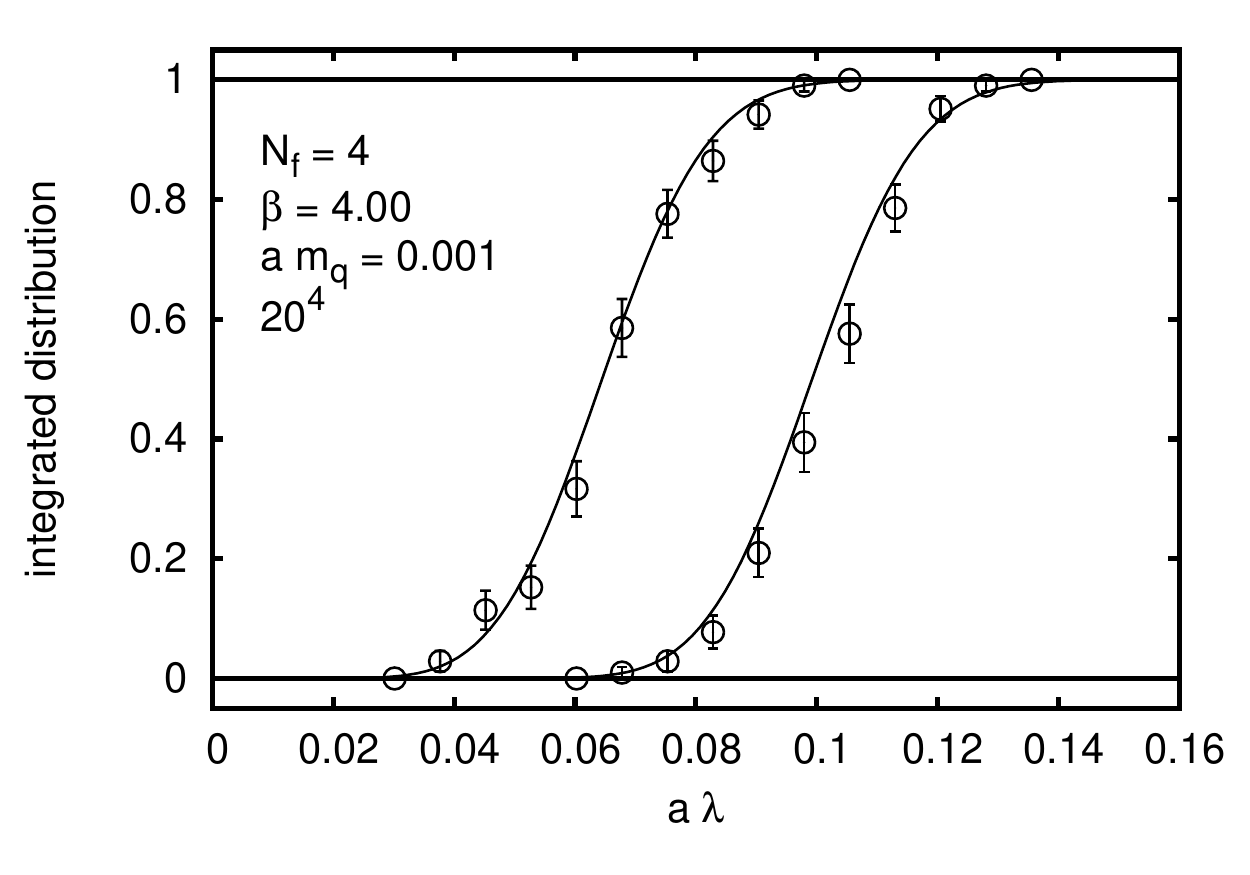}
\end{tabular}
\end{center}
\caption{From simulations at $N_f=4$ the first row shows the approach to quartet degeneracy 
of the spectrum as $\beta$ increases.
The second row shows the integrated distribution of the two lowest 
quartets averaged. The solid line compares this procedure to RMT with $N_f=4$.}   
\label{fig:rmt}
\vskip -0.1in
\end{figure*}
\section{Epsilon regime, Dirac spectrum and RMT}

If the bare parameters of a gauge theory
are tuned to the $\varepsilon$-regime in the chirally broken phase, 
the low-lying Dirac spectrum 
follows the predictions of  random matrix theory.
The corresponding random matrix model is only sensitive to the pattern of chiral symmetry breaking, 
the topological charge and the rescaled fermion mass
once the eigenvalues are also rescaled by the same factor $\Sigma_{cond} V$. 
This idea has been confirmed in various settings both
in quenched and fully dynamical simulations. The same method is applied here to nearly conformal gauge models.

The connection between the eigenvalues $\lambda$ of the Dirac operator and
chiral symmetry breaking is  given in the Banks-Casher
relation~\cite{Banks:1979yr},
$$
\Sigma_{cond} = - \langle \bar{\Psi} \Psi \rangle = \lim_{\lambda \rightarrow
  0} \lim_{m \rightarrow 0} \lim_{V \rightarrow \infty} \frac{\pi
  \rho(\lambda)}{V},
$$
where $\Sigma_{cond}$ designates the quark condensate normalized to a single flavor.
To generate a non-zero density $\rho(0)$, the smallest eigenvalues
must become densely packed as the volume increases, with an
eigenvalue spacing $\Delta \lambda \approx 1/\rho(0) = \pi/(\Sigma_{cond}
V)$. This allows a crude estimate of the quark condensate
$\Sigma_{cond}$. One can do better by exploring the $\epsilon$-regime:
If chiral symmetry is spontaneously broken, tune the volume and
quark mass such that
$
\frac{1}{F_\pi} \ll L \ll \frac{1}{M_\pi},
$
so that the pion is much lighter than the physical value, and
finite-volume effects are dominant as we discussed in Section 2. The chiral
Lagrangian of Eq.~(\ref{eq:Lagrange})
is dominated by the zero-momentum mode from the mass term and all
kinetic terms are suppressed. In this limit, the distributions of the
lowest eigenvalues are identical to those of random matrix theory, 
a theory of large matrices obeying certain symmetries~\cite{Shuryak:1992pi,Damgaard:2003nm,Verbaarschot:2000dy}. 
To connect with RMT, the eigenvalues and quark
mass are rescaled as $z = \lambda \Sigma_{cond} V$ and $\mu = m _q\Sigma_{cond} V$,
and the eigenvalue distributions also depend on the topological charge $\nu$
and the number of quark flavors $N_f$. RMT is a very useful tool to
calculate analytically all of the eigenvalue distributions. The
eigenvalue distributions in various topological sectors are measured
via lattice simulations, and via comparison with RMT, the value of the
condensate $\Sigma_{cond}$ can be extracted. 

After we generate
large thermalized ensembles, we calculate the lowest twenty eigenvalues of the
Dirac operator using the PRIMME package~\cite{primme}. In the
continuum limit, the staggered eigenvalues form degenerate quartets,
with restored taste symmetry. The first row of  Fig.~\ref{fig:rmt} shows
the change in the eigenvalue structure for $N_f=4$ as the coupling constant is varied.
At $\beta=3.6$ grouping into quartets is not seen, the pions are noticeably split, and staggered perturbation
theory is just beginning to kick in. At $\beta=3.8$ doublet pairing appears and at $\beta=4.0$ the quartets are
nearly degenerate. The Dirac spectrum is collapsed as required by the Banks-Casher 
relation.
In the second row we show the integrated distributions of the two
lowest eigenvalue quartet averages,
\begin{equation}
\int_0^{\lambda} p_k(\lambda') d\lambda', \hspace{0.5cm} k=1,2
\end{equation} 
which is only justified close to quartet degeneracy. All low eigenvalues are
selected with zero topology.  To compare with RMT, we vary $\mu=m _q\Sigma_{cond} V$ until we
satisfy 
\begin{equation}
\frac{\langle \lambda_1 \rangle_{\rm sim}}{m} = \frac{\langle z_1
  \rangle_{\rm RMT}}{\mu},
\label{eq:rmt}
\end{equation}
where $\langle \lambda_1 \rangle_{\rm sim}$ is the lowest quartet average
from simulations and the RMT average $\langle z \rangle_{\rm RMT}$
depends implicitly on $\mu$ and $N_f$. With this optimal value of
$\mu$, we can predict the shapes of $p_k(\lambda)$ and their integrated
distributions, and compare to the simulations. The agreement with the 
two lowest integrated RMT eigenvalue shapes is excellent for the larger $\beta$ values.

\begin{figure}[h!]
\begin{center}
\includegraphics[height=4.5cm]{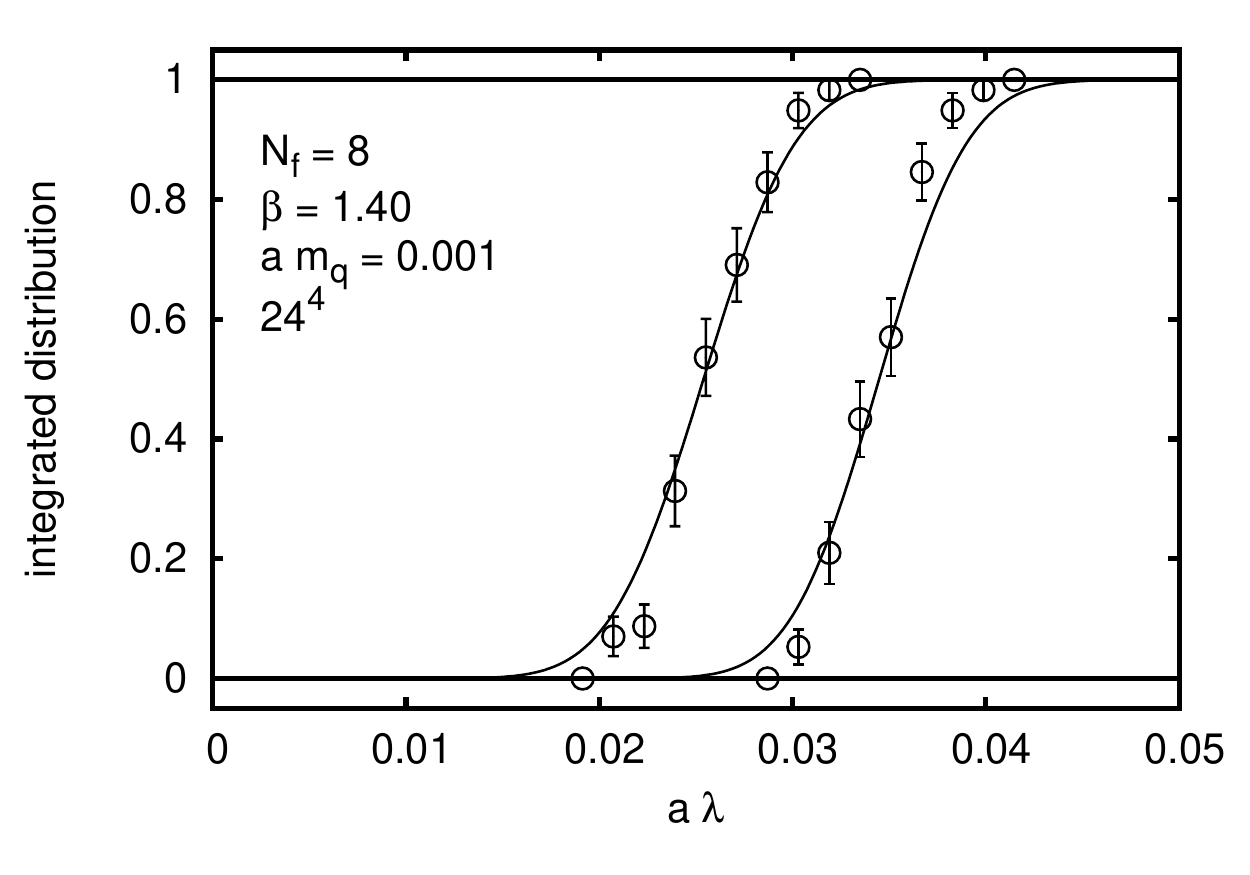}
\end{center}
\vskip -0.2in
\caption{The solid lines compare the integrated distribution of the two lowest 
quartet averages to RMT predictions with $N_f=8$.}   
\label{fig:rmtNf8}
\vskip -0.1in
\end{figure}

The main qualitative features of the RMT spectrum are very similar in our $N_f=8$ simulations as shown
in Fig.~\ref{fig:rmtNf8}. One
marked quantitative difference is a noticeable slowdown in response to change in the coupling constant.
As $\beta$ grows the recovery
of the quartet degeneracy is considerably delayed in comparison with the onset of p-regime 
Goldstone dynamics. Overall, for the $N_f=4,8$ models we find consistency between the p-regime analysis and the RMT tests.
Earlier, using Asqtad fermions at a particular $\beta$ value, we found agreement with RMT even
at $N_f=12$ which indicated a chirally broken phase~\cite{Fodor:2008hn}. Strong taste breaking with Asqtad
fermion leaves the quartet averaging in question and the bulk pronounced crossover of the Asqtad action as $\beta$
grows is also an issue. Currently we are investigating the RMT picture for $N_f=9,10,11,12$ with our much improved 
action with four and six stout steps. This action shows no artifact transitions and handles taste breaking much more
effectively. Firm conclusions on the $N_f=12$ model will require continued investigations.

\section{Inside the conformal window}

We start our investigation and simulations of the conformal window at $N_f=16$ which is the most accessible for 
analytic methods. We are particularly interested in the qualitative behavior of the finite volume spectrum of the model
and the running coupling with its associated beta function which is expected to have a weak coupling fixed
point around $g^{*2} \approx 0.5$, as estimated from the scheme independent two-loop beta function~\cite{Heller:1997vh}. 

\subsection{Conformal dynamics in finite volume}

A distinguished feature of the $N_f=16$ conformal model is how the renormalized coupling
$g^2(L)$  runs with $L$, the linear size of the spatial volume in a Hamiltonian or Transfer Matrix description. 
On very small scales the running coupling
$g^2(L)$ grows with $L$ as in any other asymptotically free theory. However, $g^2(L)$ will not grow large, and
in the $L\rightarrow \infty$ limit it will converge to the fixed
point $g^{*2}$ which is rather weak, within the reach of perturbation theory. There is
nontrivial small volume dynamics which is illustrated first in the pure gauge sector. 

At small $g^2$, without fermions,  the zero momentum components of the gauge field 
are known to dominate the dynamics~\cite{'tHooft:1979uj,Luscher:1982ma,vanBaal:1986ag}.  With $SU(3)$
gauge group, there are twenty seven degenerate vacuum states, separated by energy barriers which are generated by the
integrated effects of the non-zero momentum components of the gauge field in the Born-Oppenheimer approximation. The lowest energy excitations of the gauge field Hamiltonian scale as $ \sim  g^{2/3}(L)/L$ 
evolving into glueball states and becoming independent of the volume as the coupling constant grows with $L$. 
Nontrivial dynamics evolves through three stages as $L$ grows. In the first regime, 
in very small boxes, tunneling
is suppressed between vacua which remain isolated. In the second regime, for larger $L$, tunneling sets in and electric flux states will not be exponentially suppressed. Both regimes 
represent small worlds with zero momentum spectra separated from higher momentum modes of the theory 
with energies on the scale
of $2\pi/L$. At large enough $L$ the gauge dynamics overcomes the energy barrier, 
and wave functions spread over the vacuum valley. This third regime  is the crossover to confinement 
where the electric fluxes collapse into thin string states wrapping around the box. 

It is likely that a conformal theory with a weak coupling fixed point at $N_f=16$ will have only the first 
two regimes which are common with QCD. Now the calculations have to include fermion loops~\cite{vanBaal:1988va,Kripfganz:1988jv}.
The vacuum structure 
in small enough volumes, for which the wave functional is sufficiently localized
around the vacuum configuration, remains calculable by adding in one loop order the quantum 
effects of the fermion field fluctuations. 
The spatially constant abelian gauge fields parametrizing the vacuum valley are 
given by $A_i(\vek x)=T^aC^a_i/L$ where $T_a$ are the (N-1) generators for the Cartan subalgebra of
$SU(N)$. For $SU(3)$, $T_1=\lambda_3/2$ and $T_2=\lambda_8/2$.
With $N_f$ flavors of
massless fermion fields the effective potential of the constant mode is given by
\beq
V_{\eff}^{\vek k}(\vek C^b)=\sum_{i>j}V(\vek C^b[\mu^{(i)}_b-\mu^{(j)}_b])
-N_f\sum_{i}V(\vek C^b\mu^{(i)}_b+\pi\vek k),\label{eq:Vquark}
\eeq
with $\vek k=\vek 0$ for periodic, or $\vek k=(1,1,1)$, for  anti-periodic 
boundary conditions on the fermion fields. The function $V(\vek C)$ is the 
one-loop effective potential for $N_f=0$ and the weight 
vectors $\mu^{(i)}$ are determined by the eigenvalues of the abelian generators. 
For  SU(3) $\mu^{(1)}=
(1,1,-2)/\sqrt{12}$ and $\mu^{(2)}=\half(1,-1,0)$. 
The correct
quantum vacuum is found at the minimum of this effective potential which is dramatically
changed by the fermion loop contributions. 
\begin{figure}[ht!]
\begin{center}
\includegraphics[width=6cm,angle=0]{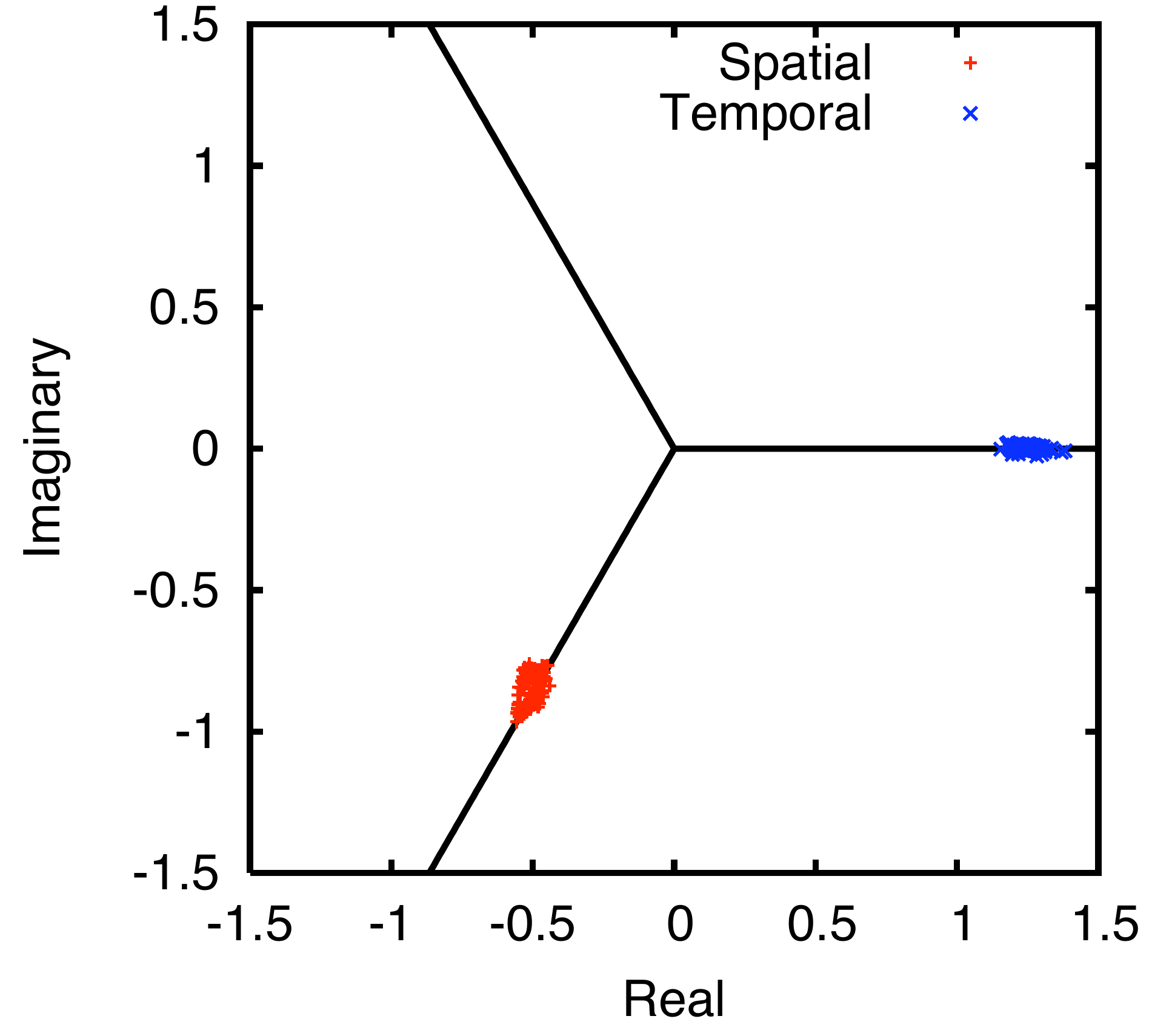}
\end{center}
\vskip -0.1in
\caption{Polyakov loop distributions, blue in the time-like and
red in the space-like directions, from our $N_f=16$ simulation with $16^4$ volume at 
$\beta=18$ with tree level Symanzik improve gauge action and staggered 
fermions with six stout steps. The fermion boundary
condition is anti-periodic in the time direction and periodic in the spatial directions.}   
\label{fig:Polyakov}
\vskip -0.1in 
\end{figure}
The  Polyakov loop observables
remain  center elements at the new vacuum configurations with complex values
\beq
P_j=\frac{1}{N}\tr\left(\exp(iC_j^bT_b)\right)=\frac{1}{N}\sum_n
\exp(i\mu^{(n)}_bC^b_j)=\exp(2\pi i l_j/N),\label{eq:PVac}
\eeq
for $SU(N)$.
This implies that $\mu^{(n)}_b\vek C^b=2\pi\vek l/N$ (mod $2\pi$), independent 
of $n$, and  $V_{\eff}^{\vek k}=-N_f NV(2\pi\vek l/N+\pi\vek k)$. In the 
case of anti-periodic boundary conditions, $\vek k=(1,1,1)$, this is minimal 
only when $\vek l=\vek 0$ (mod $2\pi$). The quantum vacuum in this 
case is the naive one, $A=0$ ($P_j=1$). In the case of periodic boundary 
conditions, $\vek k=\vek 0$, the  vacua have $\vek l\neq\vek 0$,
so that $P_j$ correspond to non-trivial center elements.
For 
SU(3),  there are now 8 degenerate vacua characterized 
by eight different Polyakov loops, $P_j=\exp(\pm 2\pi i/3)$.
Since they are related by coordinate reflections, in a small volume
parity (P) and charge conjugation (C) are spontaneously broken, although CP is 
still a good symmetry~\cite{vanBaal:1988va}. 
As shown in Fig.~\ref{fig:Polyakov}, our simulations in the $N_f=16$ model near the
fixed point $g^{*2}$ confirm this picture. In the weak coupling phase of the conformal
window the time-like Polyakov loop takes the real root, while the space-like Polyakov
loops always take the two other complex values, as expected on the basis of the above picture.
Next we will describe our method to probe the running coupling inside the conformal window.
It is a pilot study for more
comprehensive investigations of weak and strong coupling conformal dynamics.

\subsection{Running coupling and beta function}
 %
%
Consider Wilson loops $W(R,T,L)$, where $R$ and $T$ are the space-like
and time-like extents of the loop, and the lattice volume is
$L^4$ (all dimensionful quantities are expressed in units of the
lattice spacing $a$). A renormalized coupling can be defined by
\be g^2(R/L,L) = 
  - \frac{R^2}{k(R/L)} \frac{\partial^2}{\partial
  R \partial T} \ln \langle W(R,T,L) \rangle \left. \right|_{T=R},
\ee
where for convenience the definition will be restricted to Wilson loops with
$T=R$, and  $\langle ... \rangle$ is the expectation value of some
quantity over the full path integral. This definition can be motivated
by perturbation theory, where the leading term is simply the bare
coupling $g_0^2$. The renormalization scheme is defined by holding
$R/L$ to some fixed value. The quantity $k(R/L)$ is a geometric factor
which can be determined by calculating the Wilson loop expectation
values in lattice perturbation theory. The role of lattice simulations
will be to measure non-perturbatively the expectation values. On the
lattice, derivatives are replaced by finite differences, so 
the renormalized coupling is defined to be 
\bea
 && g^2((R+1/2)/L,L) = 
\frac{1}{k(R/L)} (R+1/2)^2 \chi(R+1/2,L)\, , \nn \\
&& \chi(R+1/2,L) = 
-\ln \left[ \frac{W(R+1,T+1,L)W(R,T,L)}{W(R+1,T,L)W(R,T+1,L)} \right]
\left. \right|_{T=R},\nn
\label{eq:g}
\eea
where $\chi$ is the Creutz ratio~\cite{Creutz:1980zw}, 
and the renormalization
scheme is defined by holding the value of $r=(R+1/2)/L$ fixed.

With this definition, the renormalized coupling $g^2$ is a function of
the lattice size $L$ and the fixed value of $r$. The coupling is
non-perturbatively defined, as the expectation values are calculated
via lattice simulations, which integrate over the full phase space of
the theory. By measuring $g^2(r,L)$ non-perturbatively for fixed $r$
and various $L$ values, the running of the renormalized coupling is
mapped out. In a QCD-like theory, $g^2$ increases with increasing $L$
as we flow in the infrared direction. In a conformal theory, $g^2$
flows towards some non-trivial infrared fixed point as $L$ increases,
whereas in a trivial theory, $g^2$ decreases with $L$. The advantage of
this method is that no other energy scale is required to find the
renormalization group flow. The renormalized coupling $g^2$ is also a
function of the bare coupling $g_0^2$, which is related to the lattice
spacing $a$. Keeping the lattice spacing fixed, the running of
$g^2(r,L)$ is affected by the lattice cut-off. The running has to be calculated
in the continuum limit, extrapolating to zero lattice spacing. A similar method was developed
independently in~\cite{Bilgici:2009kh}.

\begin{figure}[ht!]
\begin{center}
\includegraphics[width=0.39\textwidth]{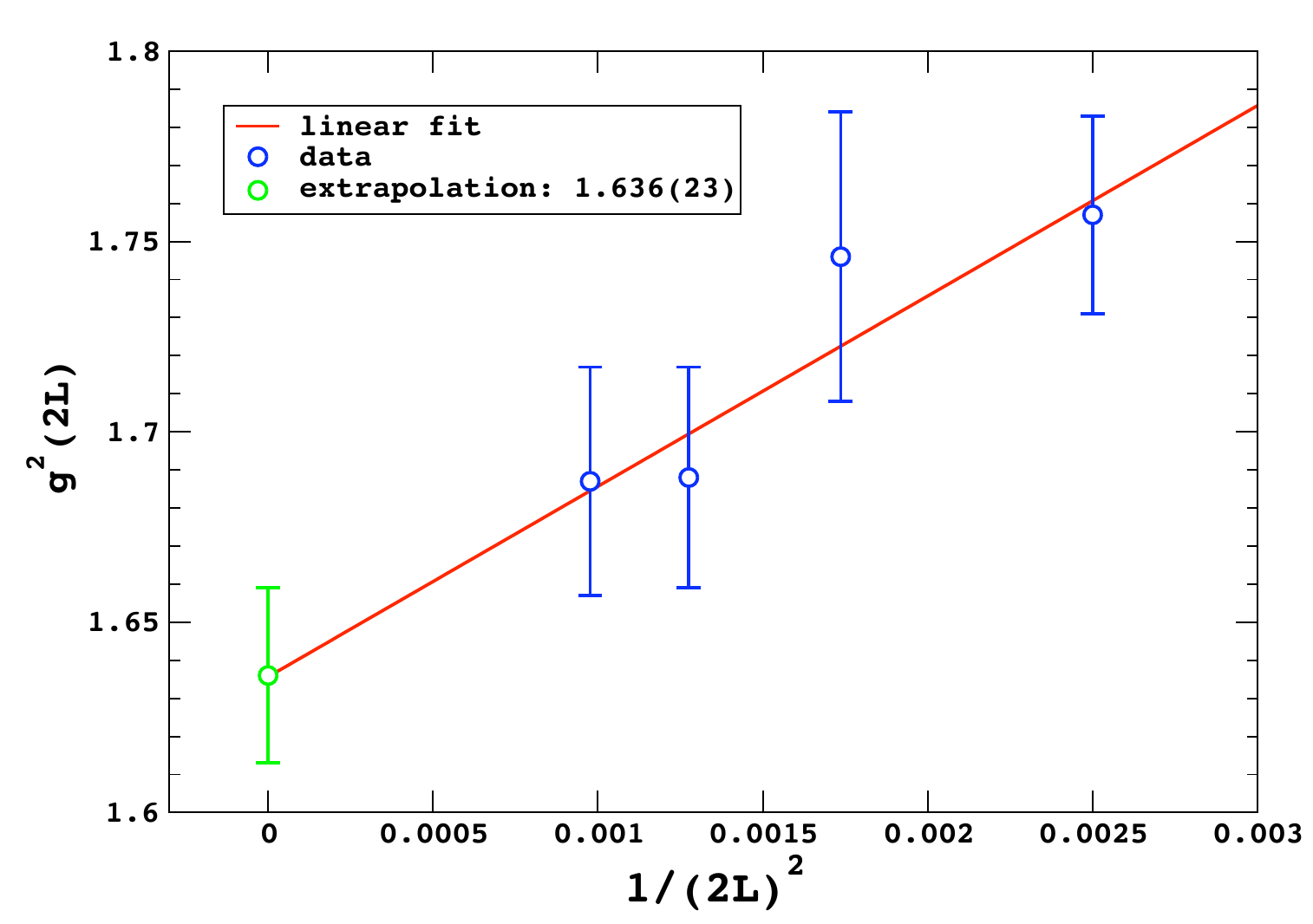}\\
\includegraphics[width=0.38\textwidth]{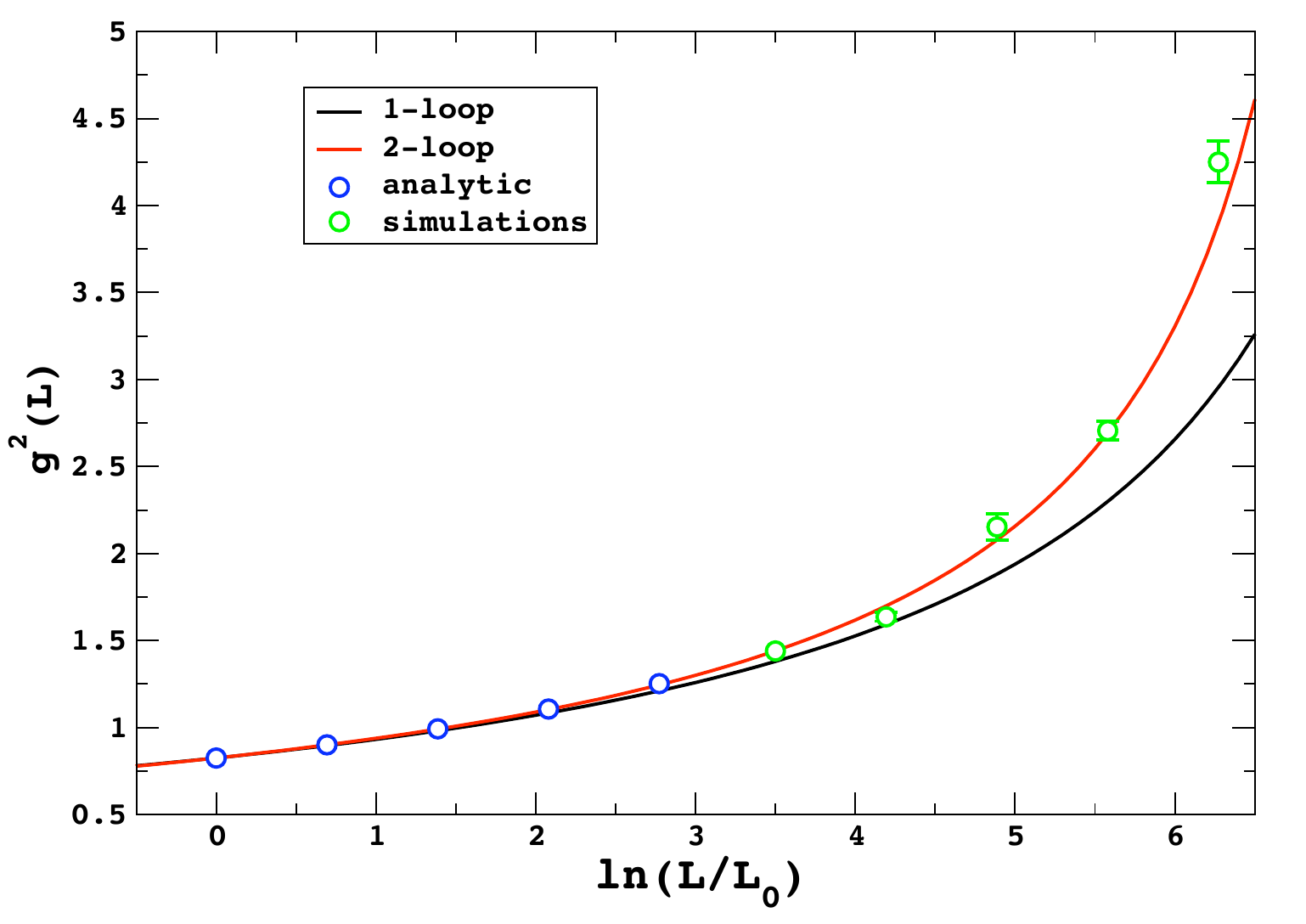}
\caption{\small The method and the main test result for pure-gauge theory 
are shown in the figure. In the upper figure the
extrapolation procedure picks up the leading $a^2/L^2$ cutoff correction term in the step
function. It gives the fit to the continuum limit value of the step function. In the lower figure,
the running coupling $g^2(L)$ is shown. The blue points are from results 
on Creutz ratios using analytic/numeric Wilson
loop lattice calculations in finite volumes with fixed value of {\em r}. In this
procedure we start from the one-loop expansion of 
Wilson loops in finite volumes based on the bare coupling~\cite{Heller:1984hx}. 
The series is re-expanded in the boosted coupling constant at the relevant
scale of the the Creutz ratio~\cite{Lepage:1992xa} to obtain realistic
estimates of our  running coupling without direct simulations. 
The rest of the procedure for the blue points
follows what we described in the text.
The green points are direct simulation results, following our procedure. The
running starts at the point $g^2(L_0)=0.825$. For almost all couplings
there is excellent agreement with continuum 2-loop running. At the
strongest coupling, the simulation results begin to break away from
perturbation theory.}
\vskip -0.3in
\label{fig:running_coupling}
\end{center}
\end{figure} 

One way to measure the running of the renormalized coupling in the
continuum limit is via step-scaling. The bare lattice
coupling is defined in the usual way $\beta = 6/g_0^2$ as it appears in the
lattice action. Some initial value of $g^2$ is picked from which
the renormalization group flow is started. On a sequence of lattice sizes
$L_1, L_2, ..., L_n$, the bare coupling is tuned on each lattice so that
exactly the same value $g^2(r,L_i,\beta_i)$ is measured from
simulations. 
Now a new set of simulations is performed, on a sequence of
lattice sizes $2L_1, 2L_2, ..., 2L_n$, using the corresponding tuned
couplings $\beta_1, \beta_2, ..., \beta_n$. From the simulations, one
measures $g^2(r,2L_i, \beta_i)$, which will vary with the bare coupling
{\it viz.}~the lattice spacing. These data can be extrapolated to the
continuum as a function of $1/L_i$. This gives one blocking step $L
\rightarrow 2L$ in the continuum renormalization group flow.
The whole procedure is then iterated. 
The chain of measurements gives the flow $g^2(r,L)
\rightarrow g^2(r,2L) \rightarrow g^2(r,4L) \rightarrow g^2(r,8L)
\rightarrow ...$, as far as is feasible~(Fig.~\ref{fig:running_coupling}). 
One is free to choose a
different blocking factor, say $L \rightarrow (3/2)L$, in which case
more blocking steps are required to cover the same energy range.

\begin{table}[ht!]
\begin{tabular}{|ccccc|}
\hline \hline
$\beta$ & $L$ &fermion mass& trajectories & $g^2(L)$\\
\hline \hline
5 & 12 &0.01& 318 & 2.06(2) \\
& 16 & 0.01&74 & 1.67(11) \\
\hline
7 & 12 &0.01& 317 & 1.207(5) \\
   & 12 & 0.001& 116 & 1.207(12)\\
& 16 &0.01& 198 & 1.13(1) \\
\hline
12 & 12&0.01 & 162 & 0.590(4) \\
& 16 &0.01& 69 & 0.577(9) \\
\hline
15 & 12&0.01 & 144 & 0.447(3) \\
&12&0.001&91 & 0.460(5)\\
& 16 &0.01& 62 & 0.444(7) \\
\hline
25 & 12 & 0.01&190 & 0.255(1) \\
& 16 &0.01& 156 & 0.253(2) \\
\hline \hline
\end{tabular}
\caption{\small Running couplings bracketing the conformal fixed point of the $N_f=16$
model in the conformal window.}
\vskip -0.1in
\end{table}


We applied the above procedure to the running coupling inside the conformal
window with $N_f=16$ flavors. The shortcut of this pilot study ignores the
extrapolation to the continuum limit. The running coupling therefore is still
contaminated with finite cutoff effects. If the linear lattice size $L$ is large
enough, the trend from the volume dependence of $g^2(L, a^2)$ should indicate 
the location of the fixed point. For $g^2(L,a^2) > g^{*2}$ we expect the decrease
of the running coupling as $L$ grows although the cutoff of the flow cannot be removed
above the fixed point. Below the fixed point with $g^2(L,a^2) < g^{*2}$
we expect the running coupling to grow as $L$ increases and the continuum limit
of the flow could be determined. The first results are summarized 
in Table 1. They are consistent with the presented picture. For example, at bare couplings
$\beta=5,7,12$ the cutoff dependent renormalized coupling is larger than $0.5$ and decreasing
with growing $L$. At small bare couplings the renormalized coupling is flat within errors and 
the flow direction is not determined. 
The independence of the results from the small quark mass of the
simulations is tested in two runs at $m_q=0.001$. 
Precise determination of the conformal fixed point in the contiuum  requires
further studies.

\vskip -0.1in
\section*{Acknowledgments}
We are thankful to Claude Bernard and Steve Sharpe for help with staggered perturbation theory and
to Ferenc Niedermayer for discussions on rotator dynamics.
We also wish to thank Urs Heller for the use of his code to calculate
Wilson loops in lattice perturbation theory, and 
Paul Mackenzie for related discussions. 
In some calculations we use the
publicly available MILC code, and the simulations were
performed on computing clusters at Fermilab, under the
auspices of USQCD and SciDAC, on the Ranger cluster of the Teragrid organization,
and on the Wuppertal GPU cluster. This work is supported by the NSF under
grant 0704171, by the DOE under grants DOE-FG03-97ER40546,
DOE-FG-02-97ER25308, by the DFG under grant FO 502/1 and by SFB-TR/55.



\end{document}